\documentstyle[12pt]{article}
\small\normalsize
\input tcilatex
\begin{document}

\author{Howard E. Brandt  \\ 
U.S. Army Research Laboratory\\2800 Powder Mill Rd., Adelphi, MD 20783 \\ }
\title{{\bf Optimum Probe Parameters for Entangling Probe in Quantum Key
Distribution}}
\date{}
\maketitle

\begin{abstract}

For the four-state protocol of quantum key distribution, optimum sets of probe parameters are calculated for the
most general unitary probe in which each individual transmitted photon is made to interact with the probe so
that the signal and the probe are left in an entangled state, and projective measurement by the probe,
made subsequent to projective measurement by the legitimate receiver, yields information about the signal state.
The probe optimization is based on maximizing the Renyi information gain by the probe on corrected data for a
given error rate induced by the probe in the legitimate receiver. An arbitrary angle is included between
the nonorthogonal linear polarization states of the signal photons. Two sets of optimum probe parameters
are determined which both correspond to the same optimization. Also, a larger set of optimum probe parameters
is found than was known previously for the standard BB84 protocol. A detailed comparison is made between
the complete and incomplete optimizations, and the latter simpler optimization is also made complete. Also,
the process of key distillation from the quantum transmission in quantum key distribution is reviewed, with
the objective of calculating the secrecy capacity of the four-state protocol in the presence of the
eavesdropping probe.  Emphasis is placed on information leakage to the probe. 

{\bf Keywords: }quantum cryptography, quantum key distribution, quantum
communication, quantum information processing
\end{abstract}

\section{INTRODUCTION}

Research efforts by many investigators have significantly advanced the field of quantum cryptography [1]
since the pioneering discoveries of Wiesner [2] and Bennett and Brassard [3,4]. Emphasis has been placed
on quantum key distribution, the generation by means of quantum mechanics of a secure random binary
sequence which can be used together with the Vernam cipher (one-time pad) [5] for secure encryption and
decryption. Various protocols have been devised for quantum key distribution, including the
single-particle four-state Bennett-Brassard protocol (BB84) [3], the single-particle two-state Bennett
protocol (B92) [6], and the two-particle entangled-state Einstein-Podolsky-Rosen (EPR) [7] protocol.
However the original BB84 protocol is presently the most practical and robust protocol.

One effective implementation of the BB84 protocol [3] uses single photons linearly polarized along one of
the four basis vectors of two sets of coplanar orthogonal bases oriented at an angle of 45 degrees
(equivalently, $ \pi /4)$ relative to each other. The polarization measurement operators in one basis do
not commute with those in the other, since they correspond to nonorthogonal polarization states. At a
fundamental level, the potential security of the key rests on the fact that nonorthogonal photon
polarization measurement operators do not commute, and this results in quantum uncertainty in the
measurement of those states by an eavesdropping probe [8]. Before transmission of each photon, the
transmitter and receiver each independently and randomly select one of the two bases. The transmitter
sends a single photon with polarization chosen at random along one of the orthogonal basis vectors in the
chosen basis. The receiver makes a polarization measurement in its chosen basis. Next, the transmitter and
the receiver, using a public communication channel, openly compare their choices of basis, without
disclosing the polarization states transmitted or received. Events in which the transmitter and the
receiver choose different bases are ignored, while the remaining events ideally have completely correlated
polarization states. The two orthogonal states in each of the two bases encode binary numbers 0 and 1, and
thus a sequence of photons transmitted in this manner can establish a random binary sequence shared by
both the transmitter and the receiver and can then serve as the secret key, following error correction and
privacy amplification [9,10]. Privacy amplification is of course necessary, because of the possibility of
an eavesdropping attack [1,3,4]. Using the Vernam cipher, the key can then be used to encode a message
which can be securely transmitted over an open communication line and then decoded, using the shared
secret key at the receiver. (The encrypted message can be created at the transmitter by adding the key to
the message and can be decrypted at the receiver by subtracting the shared secret key.)

Numerous analyses of various eavesdropping strategies have appeared in the literature. A recent review is
given in [1]. The present work is limited to an individual attack in which each transmitted photon is
measured by an independent probe after the photon polarization basis is revealed. In addition to the
individual attack, other approaches include: coherent collective attacks in which the eavesdropper
entangles a separate probe with each transmitted photon and measures all probes together as one system;
and also coherent joint attacks in which a single probe is entangled with the entire set of carrier
photons. However, these approaches require maintenance of coherent superpositions of large numbers of
states, and this is not currently feasible

For the standard four-state (BB84) protocol [3] of key distribution in quantum cryptography, Slutsky, Rao,
Sun, and Fainman [11] performed an eavesdropping probe optimization, which on average yields the most
information to the eavesdropper for a given error rate caused by the probe. The most general possible
probe consistent with unitarity was considered [11--15], in which each individual transmitted bit is made
to interact with the probe so that the carrier and the probe are left in an entangled state, and
measurement by the probe, made subsequent to measurement by the legitimate receiver, yields information
about the carrier state. The probe optimization is based on maximizing the Renyi information gain by the
probe on corrected data for a given error rate induced by the probe in the legitimate receiver.
A minimum overlap of the probe states which are correlated with the signal states
(because of the entanglement) determines the maximum Renyi information gain by the
probe. This is related to the idea that the more nearly orthogonal the correlated
probe states are, the easier they are to distinguish. The upper bound an Renyi
information gain by the probe is needed to calculate the secrecy capacity and to
determine the number of bits which must be sacrificed during privacy amplification
in order that it be exponentially unlikely that more than token leakage of the final key be available to the
eavesdropper following key distillation. The results in [11] were obtained for the standard protocol with an angle
of 45 degrees between the signal bases. The present work generalizes the probe optimization for arbitrary angle
between the signal bases.

In Section 2, a detailed review is given of the optimization of the standard BB84 protocol by Slutsky et
al [11]. In Section 3, the necessary conditions are obtained for the existence of possible extrema of the
overlap of correlated probe states for an arbitrary angle between the signal bases. Section 4 identifies
the possible extrema and associated probe parameters, and two sets of optimum probe parameters are determined
which both correspond to the optimization. Section 5 determines an analytical algebraic expression for the maximum
Renyi information gain by the probe for fixed error rate and angle between the signal bases. In
Section 6, the simplified approach of Slutsky et al, which leads to the incomplete
optimization, is made complete by recognizing certain necessary restrictions which
were ignored by those authors. In Section 7, following a review of the process of
key distillation, the secrecy capacity of  the four-state protocol in the presence
of the individual attack is calculated. Section 8 contains a summary. (The present
work reviews the results of Refs. [12--14] by the author.)

\section{PROBE OPTIMIZATION FOR STANDARD BB84 PROTOCOL}

In this Section, the probe optimization of [11] is addressed for the standard BB84 protocol in which the
angle between the signal bases is restricted precisely to $\pi /4$ (equivalently, $\alpha =\pi /8$ in
Fig.\thinspace 2 of [11]). From Section IV and Table II of [11], one has for the induced error rate $E$ in
the receiver by the eavesdropping probe, 
\begin{equation}
\label{eq}E=\frac{P_{u\stackrel{\_}{u}}+P_{\stackrel{\_}{u}u}}{P_{u\stackrel{
\_}{u}}+P_{\stackrel{\_}{u}u}+P_{uu}+P_{\stackrel{\_}{u}\stackrel{\_}{u}}},
\end{equation}
where $P_{ij}$ is the probability that if a photon in polarization state $|i> $ is transmitted in the
presence of the disturbing probe, the polarization state $|j>$ is detected by the legitimate receiver,
where $\{i,j\}=\{u, \stackrel{\_}{u},v,\stackrel{\_}{v}\}$ corresponds to nonorthogonal polarization
states $|u>$ and $|v>$, and the state $|\stackrel{\_}{u}>$ orthogonal to $|u>$, and $|\stackrel{\_}{v}>$
orthogonal to $|v>$. The states $|u>$ and $|v>$ both correspond to Boolean state $|1>$, and $|
\stackrel{\_}{u}>$ and $|\stackrel{\_}{v}>$ correspond to Boolean state $|0>$. 

One has 
\begin{equation}
\label{eq}P_{ij}=\left\langle \psi _{ij}|\psi _{ij}\right\rangle =\left|
\psi _{ij}\right| ^2,
\end{equation}
where $\left| \psi _{ij}\right\rangle $ is the projected state of the probe when polarization state
$\left| i\right\rangle $ is transmitted, and polarization state $\left| j\right\rangle $ is detected by
the receiver in the presence of the probe [11]. 

From Eqs.\thinspace (1) and (8) of [11], it follows that 
\begin{equation}
\label{eq}\left| \psi _{u\stackrel{\_}{u}}\right\rangle =\left\langle 
\stackrel{\_}{u}\right| U\left| u\otimes w\right\rangle ,
\end{equation}
where $U$ is the unitary operator producing the entanglement of the probe state $\left| w\right\rangle $
with the signal states, or 
\begin{equation}
\label{eq}\left| \psi _{u\stackrel{\_}{u}}\right\rangle =\left(
-\left\langle e_0\right| \sin \alpha +\left\langle e_1\right| \cos \alpha
\right) U\left( \left| e_0\right\rangle \cos \alpha +\left| e_1\right\rangle
\sin \alpha \right) \otimes \left| w\right\rangle ,
\end{equation}
where $\left| e_0\right\rangle $ and $\left| e_1\right\rangle $ are orthogonal basis vectors in the plane
of the polarization states of the signal, $\left| w\right\rangle $ is the initial state of the probe, and
$\alpha =\frac 12\left( \frac \pi 2-\stackrel{\_}{\theta }\right) $ is half the complement of the angle
$\stackrel{\_}{\theta }=\cos ^{-1}\left(<u|v>/|u||v|\right) $ between the two nonorthogonal
linear-polarization states $\left| u\right\rangle $ and $\left| v\right\rangle $ of the signal (see Fig. 2
of [11]; I also refer to $\stackrel{\_}{\theta }$ as the angle between the two orthogonal bases $\{\left|
u\right\rangle ,\left| \stackrel{ \_}{u}\right\rangle \}$ and $\{\left| v\right\rangle ,\left|
\stackrel{\_}{v} \right\rangle \}$.) Using Eq.\thinspace (2) of [11] in Eq.\thinspace (4), one obtains  
\begin{equation}
\label{eq}\left| \psi _{u\stackrel{\_}{u}}\right\rangle =\left(
-\left\langle e_0\right| \sin \alpha +\left\langle e_1\right| \cos \alpha
\right) \left( \cos \alpha \stackunder{n}{\sum }\left| e_n\right\rangle
\otimes \left| \Phi _{0n}\right\rangle +\sin \alpha \stackunder{n}{\sum }
\left| e_n\right\rangle \otimes \left| \Phi _{1n}\right\rangle \right) ,
\end{equation}
where $\left| \Phi _{mn}\right\rangle $ are the unnormalized nonorthogonal states of the probe. Equation
(5) becomes  
\begin{equation}
\label{eq}\left| \psi _{u\stackrel{\_}{u}}\right\rangle =\left| \Phi
_{01}\right\rangle \cos ^2\alpha -\left| \Phi _{10}\right\rangle \sin
^2\alpha +\left( \left| \Phi _{11}\right\rangle -\left| \Phi
_{00}\right\rangle \right) \sin \alpha \cos \alpha ,
\end{equation}
and substituting Eq.\thinspace (6) in Eq.\thinspace (2), and using the symmetry properties of the probe
states [11,15,16], and Eqs.\thinspace (3a), (3b), and (12) of [11], one obtains  
\begin{equation}
\label{eq}P_{u\stackrel{\_}{u}}=\frac 12(1-d)+\frac 12(d-a)\sin ^22\alpha
-\frac 12c\sin 2\alpha .
\end{equation}
where $a,b,c,$and $d$, expressed in terms of the eavesdropping probe parameters $\lambda $, $\mu $,
$\theta $, and $\phi $, are given by [11,15,16]  
\begin{equation}
\label{eq}a=\sin ^2\lambda \sin 2\mu +\cos ^2\lambda \cos 2\theta \sin 2\phi
,
\end{equation}
\begin{equation}
\label{eq}b=\sin ^2\lambda \sin 2\mu +\cos ^2\lambda \sin 2\phi ,
\end{equation}
\begin{equation}
\label{eq}c=\cos ^2\lambda \sin 2\theta \cos 2\phi ,
\end{equation}
\begin{equation}
\label{eq}d=\sin ^2\lambda +\cos ^2\lambda \cos 2\theta ,
\end{equation}
Summarizing Eq.\thinspace (7) , along with other results in Appendix C of [11], one has
\begin{equation}
\label{eq}P_{uu}=\frac 12(1+d)-\frac 12(d-a)\sin ^22\alpha +\frac 12c\sin
2\alpha ,
\end{equation}
\begin{equation}
\label{eq}P_{u\stackrel{\_}{u}}=\frac 12(1-d)+\frac 12(d-a)\sin ^22\alpha
-\frac 12c\sin 2\alpha ,
\end{equation}
\begin{equation}
\label{eq}P_{\stackrel{\_}{u}u}=\frac 12(1-d)+\frac 12(d-a)\sin ^22\alpha
+\frac 12c\sin 2\alpha ,
\end{equation}
\begin{equation}
\label{eq}P_{\stackrel{\_}{u}\stackrel{\_}{u}}=\frac 12(1+d)-\frac
12(d-a)\sin ^22\alpha -\frac 12c\sin 2\alpha ,
\end{equation}
Substituting Eqs.\thinspace (12)--(15) in Eq.\thinspace (1), one obtains 
\begin{equation}
\label{eq}E=\frac 12\left[ 1-d+(d-a)\sin ^22\alpha \right] .
\end{equation}
Also from Section IV of [11], one has for the overlap $Q$ of the probe states correlated with the signal
received by the legitimate receiver:  
\begin{equation}
\label{eq}Q=\frac{\left\langle \psi _{uu}|\psi _{\stackrel{\_}{u}\stackrel{\_
}{u}}\right\rangle }{|\psi _{uu}||\psi _{\stackrel{\_}{u}\stackrel{\_}{u}}|},
\end{equation}
or equivalently, using Eqs.\thinspace (2) in Eq.\thinspace (17), one obtains 
\begin{equation}
\label{eq}Q=\frac{\left\langle \psi _{uu}|\psi _{\stackrel{\_}{u}\stackrel{\_
}{u}}\right\rangle }{\left( P_{uu}P_{\stackrel{\_}{u}\stackrel{\_}{u}
}\right) ^{1/2}}.
\end{equation}
From Appendix C of [11], one also has 
\begin{equation}
\label{eq}\left| \psi _{uu}\right\rangle =\left| \Phi _{00}\right\rangle
\cos ^2\alpha +\left| \Phi _{11}\right\rangle \sin ^2\alpha +\left( \left|
\Phi _{10}\right\rangle +\left| \Phi _{01}\right\rangle \right) \sin \alpha
\cos \alpha ,
\end{equation}
and 
\begin{equation}
\label{eq}\left| \psi _{\stackrel{\_}{u}\stackrel{\_}{u}}\right\rangle
=\left| \Phi _{11}\right\rangle \cos ^2\alpha +\left| \Phi
_{00}\right\rangle \sin ^2\alpha -\left( \left| \Phi _{10}\right\rangle
+\left| \Phi _{01}\right\rangle \right) \sin \alpha \cos \alpha .
\end{equation}
Using Eqs.\thinspace (19), (20), the symmetry properties [11,15,16] of the probe states $\left| \Phi
_{ij}\right\rangle $, and Eqs.\thinspace (12), (3a), (3b) of [11], one obtains  
\begin{equation}
\label{eq}\left\langle \psi _{uu}|\psi _{\stackrel{\_}{u}\stackrel{\_}{u}
}\right\rangle =\frac 12\left( a+b\right) +\frac 12(d-a)\sin ^22\alpha .
\end{equation}
Next, substituting Eq.\thinspace (21), (12) and (15) in Eq.\thinspace (18), one obtains 
\begin{equation}
\label{eq}
\begin{array}{c}
Q=\left[ \frac 12\left( a+b\right) +\left( d-a\right) \frac 12\sin ^22\alpha
\right] \left[ \frac 12\left( 1+d\right) +(-d+a)\frac 12\sin ^22\alpha
+c\frac 12\sin 2\alpha \right] ^{-\frac 12} \\ 
\times \left[ \frac 12\left( 1+d\right) +(-d+a)\frac 12\sin ^22\alpha
-c\frac 12\sin 2\alpha \right] ^{-\frac 12},
\end{array}
\end{equation}
in agreement with Eq.\thinspace (15) of [11]. The optimum information gain $I_{opt}^R$ by the probe is
given in terms of the overlap $Q$ of correlated probe states by  
\begin{equation}
\label{eq}I_{\text{opt}}^R=\log _2\left( 2-Q^2\right) 
\end{equation}
(for the BB84 protocol, as well as the B92 protocol) [11,16--18]. It follows that $I_{opt}^R$ is maximized
when $Q$ is minimized. 

 It is of interest to first limit the analysis to the standard BB84 protocol in which $\alpha =\pi /8$,
corresponding to a 45-degree angle ($\stackrel{\_ }{\theta }=\pi /2-2\alpha =\pi /4$) between the signal
bases and also between the two nonorthogonal polarization states $\left| u\right\rangle $ and $\left|
v\right\rangle $ of the signal$,\;$namely, $\left\langle u|v\right\rangle =\cos \stackrel{\_}{\theta
}=\cos \left( \frac \pi 2-2\alpha \right) $= $\sin 2\alpha =\cos \left( \frac \pi 4\right) =2^{-1/2}.$ The
conditional optimization in [11] is limited to this case. In that case, Eqs.\thinspace (16) and (22)
become  
\begin{equation}
\label{eq}E_0\equiv E_{|\alpha =\pi /8}=\frac 12\left[ 1-\frac 12\left(
d+a\right) \right] ,
\end{equation}
and 
\begin{equation}
\label{eq}Q_0\equiv Q_{|\alpha =\pi /8}=\frac{\frac 12\left( d+a\right) +b}{
\left\{ \left[ 1+\frac 12\left( d+a\right) \right] ^2-\frac 12c^2\right\}
^{1/2}},
\end{equation}
respectively, in agreement with Eqs.\thinspace (15) of [11]. Substituting Eq.(24) in Eq.\thinspace (25),
the latter becomes  
\begin{equation}
\label{eq}Q_0=\frac{1-2E_0+b}{\left[ \left( 2-2E_0\right) ^2-\frac
12c^2\right] ^{1/2}},
\end{equation}
also in agreement with Eq.\thinspace (15) of [11].

For any value of $E_0$, the numerator of Eq.\thinspace (26) has a conditional (fixed $E_0$) minimum at
some point where the denominator has a conditional maximum, namely, $c=0.$ (This is further substantiated
in the following.) Clearly, the numerator of Eq.\thinspace (26) for fixed $E_0$ is minimum when $b$ is
minimum. Before minimizing $b$, substituting Eqs.\thinspace (8) and (11) in Eq.\thinspace (24), one
obtains  
\begin{equation}
\label{eq}E_0=\frac 12-\frac 14\left[ \sin ^2\lambda \left( 1+\sin 2\mu
\right) +\cos ^2\lambda \cos 2\theta \left( 1+\sin 2\phi \right) \right] , 
\end{equation}
or 
\begin{equation}
\label{eq}\sin 2\phi =\frac{2-4E_0-\sin ^2\lambda \left( 1+\sin 2\mu \right) 
}{\cos ^2\lambda \cos 2\theta }-1. 
\end{equation}
Next substituting Eq.\thinspace (28) in Eq.\thinspace (9), in order to eliminate the variable $\phi $, one
gets  
\begin{equation}
\label{eq}b=\sin ^2\lambda \sin 2\mu +\frac{2-4E_0-\sin ^2\lambda \left(
1+\sin 2\mu \right) }{\cos 2\theta }-\cos ^2\lambda . 
\end{equation}
In order that $b$ be minimum, so that $Q_0$ can be minimum in Eq.\thinspace (26), one requires that $b$ in
Eq.\thinspace (29) satisfy  
\begin{equation}
\label{eq}\frac{\partial b}{\partial \mu }=0, 
\end{equation}
\begin{equation}
\label{eq}\frac{\partial b}{\partial \lambda }=0, 
\end{equation}
and 
\begin{equation}
\label{eq}\frac{\partial b}{\partial \theta }=0. 
\end{equation}
Substituting Eq.\thinspace (29) in Eqs.\thinspace (30), (31), and (32), one requires 
\begin{equation}
\label{eq}\sin ^2\lambda \cos 2\mu \left( 1-\frac 1{\cos 2\theta }\right)
=0, 
\end{equation}
\begin{equation}
\label{eq}\sin 2\lambda \left( \sin 2\mu +1\right) \left( 1-\frac 1{\cos 
2\theta }\right) =0, 
\end{equation}
\begin{equation}
\label{eq}\frac{\sin 2\theta }{\cos ^22\theta }\left[ 2-4E_0-\sin ^2\lambda
\left( 1+\sin 2\mu \right) \right] =0. 
\end{equation}
Equations (33)--(35) are necessary conditions for minimum $b$ and $Q_{0\text{
.}}$

Equation (33) requires 
\begin{equation}
\label{eq}\text{(ia)\ \ \ \ \ \ \ \ \ }\sin \lambda =0 
\end{equation}
or 
\begin{equation}
\label{eq}\;\;\text{(ib)\ \ \ \ \ \ \ \ \ cos2}\mu =0 
\end{equation}
or 
\begin{equation}
\label{eq}\;\;\text{(ic)\ \ \ \ \ \ \ \ \ cos2}\theta =1. 
\end{equation}
Equation (34) requires 
\begin{equation}
\label{eq}\;\text{(iia)\ \ \ \ \ \ \ \ }\sin 2\lambda =0 
\end{equation}
or 
\begin{equation}
\label{eq}\;\text{(iib)\ \ \ \ \ \ \ \ }\sin 2\mu =-1 
\end{equation}
or 
\begin{equation}
\label{eq}\;\;\text{(iic)\ \ \ \ \ \ \ \ \ }\cos 2\theta =1. 
\end{equation}
Equation (35) requires 
\begin{equation}
\label{eq}\text{(iiia)\ \ \ \ \ \ \ \ \ \ \ \ }\sin 2\theta =0 
\end{equation}
or 
\begin{equation}
\label{eq}\;\;\;\;\;\;\;\;\;\;\;\;\;\;\;\;\;\;\;\;\;\;\;\;\;\;\;\text{
(iiib)\ \ \ \ \ \ \ \ \ \ \ \ \ }\sin ^2\lambda \left( 1+\sin 2\mu \right)
=2-4E_0. 
\end{equation}

A solution to Eqs.\thinspace (33)--(35), which leads to the optimization given in [11], is given by 
\begin{equation}
\label{eq}\sin \lambda =0;\;\;\;\;\sin 2\theta =0;\;\;\;\;\cos 2\theta
=e_\theta \equiv \pm 1.
\end{equation}
Equations (44) satisfy Eqs.\thinspace (ia), (iia), and (iiia), and therefore also Eqs.\thinspace
(33)--(35). Next, substituting Eqs.\thinspace (44) in Eq.\thinspace (10), one gets  
\begin{equation}
\label{eq}c=0,
\end{equation}
consistent with the conditional maximum of the denominator in Eq.\thinspace (26), as declared above.

Furthermore, substituting Eqs.\thinspace (44) in Eq.\thinspace (28), one obtains 
\begin{equation}
\label{eq}\sin 2\phi =\frac 2{e_\theta }\left( 1-2E_0\right) -1.
\end{equation}
Since only $E_0<1/2$ is considered [11], and clearly $E_0\geq 0$, then one requires 
\begin{equation}
\label{eq}0\leq E_0<1/2.
\end{equation}
Then substituting Eq.\thinspace (46) in Eq.\thinspace (47), one requires 
\begin{equation}
\label{eq}0<e_\theta \left( \sin 2\phi +1\right) \leq 2.
\end{equation}
Clearly one requires $e_\theta =+1$ because if $e_\theta =-1$, then Eq.\thinspace (48) implies $\sin
2\varphi <-1$, which is impossible. Therefore, one has in Eq.\thinspace (44),  
\begin{equation}
\label{eq}\cos 2\theta =e_\theta =1,
\end{equation}
and Eq.\thinspace (48) becomes 
\begin{equation}
\label{eq}-1<\sin 2\phi \leq 1.
\end{equation}

Next substituting Eqs.\thinspace (44) and (49) in Eqs.\thinspace (8)--(11), one requires 
\begin{equation}
\label{eq}a=\sin 2\phi ,
\end{equation}
\begin{equation}
\label{eq}b=\sin 2\phi ,
\end{equation}
\begin{equation}
\label{eq}c=0,
\end{equation}
and 
\begin{equation}
\label{eq}d=1.
\end{equation}
(Equation (53) restates Eq.\thinspace (45).) Next substituting Eqs.\thinspace (51) and (54) in
Eq.\thinspace (24), one obtains  
\begin{equation}
\label{eq}E_0=\frac 14\left( 1-\sin 2\phi \right) ,
\end{equation}
and therefore 
\begin{equation}
\label{eq}\sin 2\varphi =1-4E_0.
\end{equation}
Also, substituting Eqs.\thinspace (52), (53) and (56) in Eq.\thinspace (26), one obtains 
\begin{equation}
\label{eq}Q_0=3-\frac 2{1-E_0}.
\end{equation}
Equations (57), (44), (49), and (50)--(55) agree with Eqs.\thinspace (16) of [2]. The choice of $\mu =0$
in [11] is allowed because $\mu $ only enters through $a$ and $b$ in Eqs.\thinspace (8) and (9), and
according to Eq.\thinspace (44), $\sin \lambda =0$. In general, however, any $\mu $ ($0\leq \mu \leq \pi
$) produces the same optimization. Also, $\lambda =\pi $ satisfies Eq.\thinspace (44) as well as $\lambda
=0$. Other combinations of Eqs.\thinspace (36)--(43) may also yield solutions, and this issue is addressed
in Section 4 for arbitrary values of $\alpha $. 

It is also well to further clarify the arguments of Appendix E in [11]. Note that according to
Eq.\thinspace (9) above, $b$ is independent of $\theta ,$ and $E_0$ in Eq.\thinspace
(27) is clearly least when
$\cos 2\theta =1,$ since in the last term of Eq.\thinspace (27), $\cos ^2\lambda \geq 0,$ and according to
Eq.\thinspace (50), $0<\left( 1+\sin 2\phi \right) \leq 2$. But then substituting Eq.\thinspace (49) in
Eq.\thinspace (27), the latter becomes  
\begin{equation}
\label{eq}E_0=\frac 12-\frac 14\left[ 1+\sin ^2\lambda \sin 2\mu +\cos
^2\lambda \sin 2\phi \right] .
\end{equation}
Substituting Eq.\thinspace (9) in Eq.\thinspace (58), then 
\begin{equation}
\label{eq}E_0=\frac 14\left[ 1-b\right] ,
\end{equation}
which agrees with Eqs.\thinspace (52) and (55). According to Eq.\thinspace (59), $E_0$ is a monotonically
decreasing function of $b$, and the problem of minimizing $b$, subject to constant $E$, can be inverted so
that $E$ is minimized, subject to constant $b$. One also sees by substituting Eqs.\thinspace (59) and (53)
in Eq.\thinspace (26) that Eq.\thinspace (57) is again obtained, and since Eq.\thinspace (57) results from
minimizing $b$ with $E_{0\text{ }}$constant, this is equivalent to minimizing $E_0$ with $b$ constant, and
is consistent with Appendix E of [11]. This approach to the optimization is further examined in section (6).

\section{Extrema and Probe Parameters}

In this Section, conditions for possible relative extrema are calculated of the overlap of correlated
probe states of the Fuchs-Peres probe [11,15] for an arbitrary angle between the signal bases. Although the
standard BB84 protocol with angle $\pi/4$ between the signal bases will be seen to yield the most information
to the probe, sensitivity to practical tuning variations in this angle can be useful in quantifying
tolerances. First, Eq.\thinspace (22) can be rewritten as  
\begin{equation}
\label{eq}Q=\frac{\frac 12\left( a+b\right) +\left( d-a\right) \frac 12\sin
^22\alpha }{\left\{ \frac 14\left[ 1+d+\left( a-d\right) \sin ^22\alpha
\right] ^2-\frac 14c^2\sin ^22\alpha \right\} ^{1/2}}.
\end{equation}
Also, from Eq.\thinspace (16), it follows that 
\begin{equation}
\label{eq}\left( d-a\right) \sin ^22\alpha =2E-1+d,
\end{equation}
and substituting Eq.\thinspace (61) in Eq.\thinspace (60), one obtains 
\begin{equation}
\label{eq}Q=\frac{\frac 12\left( a+b+d-1\right) +E}{\left\{ \left(
1-E\right) ^2-\frac 14c^2\sin ^22\alpha \right\} ^{1/2}}.
\end{equation}

From Eq.\thinspace (61), it follows that 
\begin{equation}
\label{eq}d=-\frac{2E-1+a\sin ^22\alpha }{\cos ^22\alpha }. 
\end{equation}
Next, using Eqs.\thinspace (8), (9), and (63), and defining a quantity $q$ to be $[a + b + d]$, one can show that 
\begin{equation}
\label{eq}
\begin{array}{c}
q\equiv a+b+d=\left( 2-\tan ^22\alpha \right) \sin ^2\lambda \sin 2\mu
\;\;\;\;\;\;\;\;\;\;\;\;\;\;\;\;\;\;\;\;\;\;\;\;\;\;\;\;\;\;\;\;\;\;\;\;\;\;
\;\;\;\; \\ 
+\cos ^2\lambda \sin 2\phi \left[ 1+\left( 1-\tan ^22\alpha \right) \cos
2\theta \right] -\frac{2E-1}{\cos ^22\alpha }. 
\end{array}
\end{equation}

Next substituting Eqs.\thinspace (8) and (11) in Eq.(16), one has 
\begin{equation}
\label{eq}
\begin{array}{c}
E=\frac 12\left[ 1-\sin ^2\lambda -\cos ^2\lambda \cos 2\theta \right.
\;\;\;\;\;\;\;\;\;\;\;\;\;\;\;\;\;\;\;\;\;\;\;\;\;\;\;\;\;\;\;\;\;\;\;\;\;\;
\;\;\;\;\;\;\;\;\;\;\;\;\;\;\;\;\;\; \\ 
\;\;\;\;\;\;\;\left. +\sin ^22\alpha \left( \sin ^2\lambda +\cos ^2\lambda
\cos 2\theta -\sin ^2\lambda \sin 2\mu -\cos ^2\lambda \cos 2\theta \sin
2\phi \right) \right] . 
\end{array}
\end{equation}
It then follows from Eq.\thinspace (65) that 
\begin{equation}
\label{eq}
\begin{array}{c}
\sin 2\mu
=\;\;\;\;\;\;\;\;\;\;\;\;\;\;\;\;\;\;\;\;\;\;\;\;\;\;\;\;\;\;\;\;\;\;\;\;\;
\;\;\;\;\;\;\;\;\;\;\;\;\;\;\;\;\;\;\;\;\;\;\;\;\;\;\;\;\;\;\;\;\;\;\;\;\;\;
\;\;\;\;\;\;\;\; \\     
\;\;\;\;\;\;\;\;\;\;\frac{\cos ^2\lambda \left( 1-\cos 2\theta \right) +\sin  
^22\alpha \left( \sin ^2\lambda +\cos ^2\lambda \cos 2\theta -\cos ^2\lambda
\cos 2\theta \sin 2\phi \right) -2E}{\sin ^22\alpha \sin ^2\lambda }. 
\end{array}
\end{equation}
Next substituting Eq.\thinspace (66) in Eq.\thinspace (64) to eliminate dependence on $\mu $, it follows
that  
\begin{equation}
\label{eq}
\begin{array}{c}
q\equiv a+b+d=\cos ^2\lambda \left\{ \left( 2-\tan ^22\alpha \right) \left[
\cot ^22\alpha -\cos 2\theta \left( \sin 2\phi +\cot ^22\alpha \right)
\right] \right. \\ 
+\left. \sin 2\phi \left[ 1+\left( 1-\tan ^22\alpha \right) \cos 2\theta
\right] \right\} -4\csc ^22\alpha \,E+3. 
\end{array}
. 
\end{equation}
Also, substituting the definition of $q$, Eq.\thinspace (64) in Eq.\thinspace (62), one obtains 
\begin{equation}
\label{eq}Q=\frac{\frac 12\left( q-1\right) +E}{\left[ \left( 1-E\right)
^2-\frac 14c^2\sin ^22\alpha \right] ^{1/2}}, 
\end{equation}
where $q$ is given by Eq.\thinspace (67), $c$ is given by Eq.\thinspace (10), and $E$ is constant. Since
$q$ and $c$ depend only on $\lambda $, $\theta $, and $\phi $, and since $E$ is constant, then $Q$ depends
only on the variables $\lambda $, $\theta $, and $\phi .$ 

Possible extrema of the overlap $Q$ for fixed $E$ must satisfy 
\begin{equation}
\label{eq}\frac{\partial Q}{\partial \lambda }=0, 
\end{equation}
\begin{equation}
\label{eq}\frac{\partial Q}{\partial \theta }=0, 
\end{equation}
\begin{equation}
\label{eq}\frac{\partial Q}{\partial \phi }=0. 
\end{equation}
In general, Eqs.\thinspace (69)--(71) may determine absolute or relative maximum, minimum, or saddle
points in the space of probe parameters. The minimum $Q$ is sought here. Possible solutions to Eqs. (69)--(71),
giving the values of the probe parameters at the possible extrema, are derived in the Appendix. Each possible
solution corresponds to one of the combinations given by Eqs. (A-39)--(A-50), in which the fractions $F_1, F_2,$
and $F_3$ are defined by Eqs. (A-5), (A-10), and (A-15), respectively. 

\section{OPTIMUM PROBE PARAMETERS}

Possible solutions to Eqs. (69)--(71) summarized in the Appendix by Eqs. (A-39)--(A-50) are designated by
possibilities (A)--(L), respectively.

Possibilities (A), (C), (D) and (J) are excluded in the Appendix. Possibilities (B), (E)--(I), (K), and (L)
all gave the same result, Eq.\thinspace (A-60). However they differ in the values of the optimized probe
parameters. 

First consider possibility (B). According to Eqs.\thinspace (A-51), (A-54), and (A-55), one has for the probe
parameters $\lambda $, $\mu $, $\theta $, and $\phi $:  
\begin{equation}
\label{eq}\sin \lambda =0, 
\end{equation}
\begin{equation}
\label{eq}\cos 2\theta =1, 
\end{equation}
\begin{equation}
\label{eq}\sin 2\phi =1-2E\csc ^22\alpha . 
\end{equation}
Evidently, according to Eqs.\thinspace (72) and (66), the probe parameter $\mu $ is arbitrary ($0\leq \mu
\leq \pi )$. In summary then for possibility (B), the optimized probe parameters are:  
\begin{equation}
\label{eq}\left\{ \lambda ,\mu ,\theta ,\phi ;\;\sin \lambda =0,\;\cos
2\theta =1,\;\sin 2\phi =1-2E\csc ^22\alpha \right\} . 
\end{equation}

Next consider possibility (E). According to Eqs.\thinspace (A-134) and (A-135), one has 
\begin{equation}
\label{eq}\cos \lambda =0, 
\end{equation}
\begin{equation}
\label{eq}\sin 2\mu =1-2E\csc ^22\alpha . 
\end{equation}
Evidently $\theta$ and $\phi$ are arbitrary ($0\leq \theta \leq \pi, 0\leq \phi \leq \pi )$. Thus for
possibility (E), the optimized probe parameters are  
\begin{equation}
\label{eq}\left\{ \lambda ,\mu ,\theta ,\phi ;\;\cos \lambda =0,\;\sin 2\mu
=1-2E\csc ^22\alpha \right\} . 
\end{equation}

For possibility (F), according to Eqs.\thinspace (A-141), (A-146), (A-143), and (A-147), the optimized probe
parameters are:  
\begin{equation}
\label{eq}\left\{ \lambda ,\mu ,\theta ,\phi ;\;\sin 2\mu \sin ^2\lambda
=1-2E\csc ^22\alpha \mp \cos ^2\lambda ,\;\cos 2\theta =1,\;\sin 2\phi =\pm
1\right\} . 
\end{equation}

For possibility (G), according to Eqs.\thinspace (A-148), (A-150), (A-152), and (A-154), the optimized probe
parameters are:  
\begin{equation}
\label{eq}\left\{ \lambda ,\mu ,\theta ,\phi ;\;\cos \lambda =0,\;\sin 2\mu
=1-2E\csc ^22\alpha ,\;\cos 2\theta =1\right\} , 
\end{equation}
or 
\begin{equation}
\label{eq}\left\{ \lambda ,\mu ,\theta ,\phi ;\;\cos \lambda =0,\;\sin 2\mu
=1-2E\csc ^22\alpha ,\;\sin 2\phi =1-2\cot ^22\alpha ,\;\cos 2\theta
=e_\theta \right\} , 
\end{equation}
Equations (80) and (81) are apparently included in Eq.\thinspace (78).

For possibility (H), according to Eqs.\thinspace (A-160) and (A-162), one has 
\begin{equation}
\label{eq}\left\{ \lambda ,\mu ,\theta ,\phi ;\;\sin 2\mu \sin ^2\lambda
=1-2E\csc ^22\alpha -\cos ^2\lambda \sin 2\phi ,\;\cos 2\theta =1\right\} . 
\end{equation}
Evidently Eqs.\thinspace (75) and (79) are included in Eq.\thinspace (82).

For possibility (I), according to Eqs.\thinspace (A-163), (A-165), (A-167), and (A-168), one has 
\begin{equation}
\label{eq}\left\{ \lambda ,\mu ,\theta ,\phi ;\;\cos \lambda =0,\;\sin 2\mu
=1-2E\csc ^22\alpha ,\;\cos 2\theta =1\right\} , 
\end{equation}
or, alternatively, 
\begin{equation}
\label{eq}\left\{ \lambda ,\mu ,\theta ,\phi ;\;\cos \lambda =0,\;\sin 2\mu
=1-2E\csc ^22\alpha ,\;\sin 2\phi =1-2\cot ^22\alpha \right\} . 
\end{equation}
Equations (83) and (84) are evidently included in Eq.\thinspace (78).

For possibility (K), according to Eqs.\thinspace (A-174), (A-177), and (A-179), the optimum probe parameters
are:  
\begin{equation}
\label{eq}\left\{ \lambda ,\mu ,\theta ,\phi ;\;\cos \lambda =0,\;\sin 2\mu
=1-2E\csc ^22\alpha ,\;\sin 2\phi =1-2\cot ^22\alpha \right\} . 
\end{equation}
Comparing Eq.\thinspace (85) with Eq.\thinspace (78), it is evident that Eq.\thinspace (85) is included
in Eq.\thinspace (78)\thinspace 

Finally, for possibility (L), according to Eqs.\thinspace (A-186), (A-188) and (A-189), the optimum probe
parameters are  
\begin{equation}
\label{eq}
\begin{array}{c}
\left\{ \lambda ,\mu ,\theta ,\phi ;\;\sin 2\mu \sin ^2\lambda =1-2E\csc
^22\alpha -\cos ^2\lambda \sin 2\phi ,\;\cos 2\theta
=1,\;\;\;\;\;\;\;\;\;\;\;\;\;\right.  \\ 
\cos ^2\lambda =2\left( 1-E\right) ^2\left( 1-2\cot ^22\alpha -\sin 2\phi
\right)  \\ 
\times \left. \left[ \sin ^22\alpha \cos ^22\phi \left[ 1+\left( 1-2\csc
^22\alpha \right) E\right] \right] ^{-1}\right\} .
\end{array}
.
\end{equation}
Comparing Eqs.\thinspace (86) with Eq.\thinspace (78), one sees that Eq.\thinspace (86) is included in
Eq.\thinspace (82). 

Equations (78) and (82) are different possible sets of optimized probe parameters, both of which
correspond to the same optimization, Eq.\thinspace (A-60). In summary, the optimized sets of probe
parameters are:  
\begin{equation}
\label{eq}\left\{ \lambda ,\mu ,\theta ,\phi ;\;\cos \lambda =0,\;\sin 2\mu
=1-2E\csc ^22\alpha \right\} , 
\end{equation}
\begin{equation}
\label{eq}\left\{ \lambda ,\mu ,\theta ,\phi ;\;\sin 2\mu \sin ^2\lambda
=1-2E\csc ^22\alpha -\cos ^2\lambda \sin 2\phi ,\;\cos 2\theta =1\right\} . 
\end{equation}

For $\alpha =\pi /8$, these reduce to 
\begin{equation}
\label{eq}\left\{ \lambda ,\mu ,\theta ,\phi ;\;\cos \lambda =0,\;\sin 2\mu
=1-4E\right\} , 
\end{equation}
\begin{equation}
\label{eq}\left\{ \lambda ,\mu ,\theta ,\phi ;\;\sin 2\mu \sin ^2\lambda
=1-4E-\cos ^2\lambda \sin 2\phi ,\;\cos 2\theta =1\right\} . 
\end{equation}
Equation (90), for $\sin \lambda =0$, corresponds to the standard optimization in [2] and in Section 2
above, but, other than that, the two sets of optimized probe parameters given by Eqs.\thinspace (89) and
(90) were not found by the simplified arguments appearing there. (Still another set of solutions, holding only for
$\alpha = \pi/4$, follows from Eq. (A-173), and is addressed in Section 6.) Both Eqs.\thinspace (89) and (90)
(together with Eqs.\thinspace (8)--(11), (24), and (26)) yield Eq.\thinspace (57). It can also be shown that all
sets of optimum probe parameters following from Eqs.\thinspace (36)--(43) are subsets of Eq.\thinspace (88), and
also yield Eq.\thinspace (57). 

\section{MAXIMUM INFORMATION GAIN}

In Section 4 and the Appendix, it was determined that the only remaining possible extremum of the overlap $Q$ of
correlated probe states for fixed error rate $E$ is given by Eq.\thinspace (A-60), namely,
\begin{equation}
Q={{1+(1-2\csc ^22\alpha )E}
\over {1-E}} .
\end{equation}
I have found that if one plots
points using the general expression for the nonoptimized overlap given by the parametric Eqs.\thinspace
(60) and (16) along with Eqs.\thinspace (8)--(11) for a representative range of values of the error rate
$E$ and the probe parameters $\lambda $, $\mu $, $\theta $, and $\phi ,$ for a range of $\alpha \leq \pi
/8$, the nonoptimized values of $Q$ all lie above the corresponding curves given by Eq.\thinspace (91).
Also, by explicitly calculating the difference between the optimized overlap, Eq.\thinspace (91), and the
nonoptimized overlap, Eqs.\thinspace (60) and (16), for representative ranges of the error rate and the
probe parameters in the neighborhood surrounding each of the optimized sets, Eqs.\thinspace (87) and
(88), I have found that for $\alpha =\pi /8$ or $\pi /9$, the nonoptimized overlap is not decreasing, and
therefore Eq.\thinspace (91) does in fact represent a minimum. Also, it is evident from Eq.\thinspace
(91) that the minimum overlap $Q,$ for constant $E$, decreases as $\alpha $ decreases below $\pi /8$.
Apparently, the optimization holds for $\alpha \leq \pi /8.$ However, for $\alpha >\pi /8$, this is not
the case (points resulting from Eqs.\thinspace (60) and (16) fall above and below the curves given by
Eq.\thinspace (91)), and therefore the extremization does not correspond to a minimum for $\alpha >\pi
/8$. (For example, if $\alpha =\pi /8+10^{-6}$, $E=0.2$, $\mu /\pi =0.156816$, $\lambda /\pi =0.3$,
$\theta /\pi =0.1$, and $\phi /\pi =0.75$, one obtains, using Eqs.\thinspace (16), (60), and (8)--(11),
the value $Q=0.500003$ for the nonoptimized overlap; but Eq.\thinspace (91) yields a larger value,
$Q=0.500004$. Also, if $\alpha =\pi /5$, $E=0.3$, $\mu /\pi =0.0711275$, $\lambda /\pi =0.7$, $\theta /\pi
=0.7$, and $\phi /\pi =0.7$, one obtains $Q=0.34828$ for the nonoptimized overlap, but Eq.\thinspace (91)
yields $Q=0.909509$.) 

However, it is at this point essential to note the invariance of the Error rate $E$, Eq.\thinspace (1),
and the overlap $Q$, Eq.\thinspace (17), under an interchange of the states $\left| u\right\rangle $ and
$\left| \stackrel{\_}{u}\right\rangle $; thus  
\begin{equation}
\label{eq}\left\{ E,Q\right\} \stackunder{\left| u\right\rangle
\leftrightarrow \left| \stackrel{\_}{u}\right\rangle }{\longrightarrow }
\left\{ E,Q\right\} .
\end{equation}
Also, from Fig.\thinspace 2 of [11], it is evident that under the interchange of $\left| u\right\rangle $
and $\left| \stackrel{\_}{u}  
\right\rangle $, the angle $\stackrel{\_}{\theta }$ between the nonorthogonal polarization states becomes
$2\alpha $; thus  
\begin{equation}
\label{eq}\stackrel{\_}{\theta }\stackunder{\left| u\right\rangle
\leftrightarrow \left| \stackrel{\_}{u}\right\rangle }{\longrightarrow }
2\alpha ,
\end{equation}
or equivalently, since $\stackrel{\_}{\theta }=\frac \pi 2-2\alpha $, 
\begin{equation}
\label{eq}\alpha \stackunder{\left| u\right\rangle \leftrightarrow \left| 
\stackrel{\_}{u}\right\rangle }{\longrightarrow }\frac \pi 4-\alpha .
\end{equation}
Also, using Eq.\thinspace (94), one has 
\begin{equation}
\label{eq}\left\{ \alpha \leq \pi /8\right\} \stackunder{\left|
u\right\rangle \leftrightarrow \left| \stackrel{\_}{u}\right\rangle }{
\longrightarrow }\left\{ \alpha \geq \pi /8\right\} .
\end{equation}
It then follows from Eqs.\thinspace (91), (94), and (95) that the optimum overlap, 
\begin{equation}
\label{eq}Q=\frac{1+\left( 1-2\csc ^22\alpha \right) E}{1-E},\;\;\alpha \leq
\pi /8,
\end{equation}
becomes 
\begin{equation}
\label{eq}Q=\frac{1+\left( 1-2\csc ^22\left( \frac \pi 4-\alpha \right)
\right) E}{1-E},\;\;\alpha \geq \pi /8,
\end{equation}
or equivalently, 
\begin{equation}
\label{eq}Q=\frac{1+\left( 1-2\sec ^22\alpha \right) E}{1-E},\;\;\alpha \geq
\pi /8.
\end{equation}
Also, the optimized sets of probe parameters, Eqs.\thinspace (87) and (88), namely, 
\begin{equation}
\label{eq}\left\{ \lambda ,\mu ,\theta ,\phi ;\;\cos \lambda =0,\;\sin 2\mu
=1-2E\csc ^22\alpha \right\} ,\;\;\alpha \leq \pi /8,
\end{equation}
\begin{equation}
\label{eq}\left\{ \lambda ,\mu ,\theta ,\phi ;\;\sin 2\mu \sin ^2\lambda
=1-2E\csc ^22\alpha -\cos ^2\lambda \sin 2\phi ,\;\cos 2\theta =1\right\}
,\;\;\alpha \leq \pi /8,
\end{equation}
become, for $\alpha \rightarrow \frac \pi 4-\alpha :$
\begin{equation}
\label{eq}\left\{ \lambda ,\mu ,\theta ,\phi ;\;\cos \lambda =0,\;\sin 2\mu
=1-2E\sec ^22\alpha \right\} ,\;\;\alpha \geq \pi /8,
\end{equation}
\begin{equation}
\label{eq}\left\{ \lambda ,\mu ,\theta ,\phi ;\;\sin 2\mu \sin ^2\lambda
=1-2E\sec ^22\alpha -\cos ^2\lambda \sin 2\phi ,\;\cos 2\theta =1\right\}
,\;\;\alpha \geq \pi /8.
\end{equation}
I have found that if one plots points using the general expression for the nonoptimized overlap, given by
the parametric Eqs.\thinspace (60) and (16) along with Eqs.\thinspace (8)--(11), for a representative
range of values of the error rate $E$ and the probe parameters $\lambda $, $\mu $, $\theta $, and $\phi ,$
for a range of $\alpha \geq \pi /8$, the nonoptimized values of $Q$ all lie above the corresponding curves
given by Eq.\thinspace (98). Apparently, for $\alpha \geq \pi /8$, the optimization, Eq.\thinspace (98),
holds. 

With the restrictions on $\alpha $, the maximum Renyi information gain by the probe is given by
Eq.\thinspace (23), namely, [11--14]  
\begin{equation}
\label{eq}I_{opt}^R=\log _2(2-Q^2),
\end{equation}
where $Q$ is given by Eq.\thinspace \thinspace (96) for $\alpha \leq \pi /8$, and Eq.(98) for $\alpha
\geq \pi /8$, or  
\begin{equation}
\label{eq}Q=\left\{ 
\begin{array}{c}
\frac{1+\left( 1-2\csc ^22\alpha \right) E}{1-E},\;\;\alpha \leq \pi /8 \\ 
\frac{1+\left( 1-2\sec ^22\alpha \right) E}{1-E},\;\;\alpha \geq \pi /8
\end{array}
\right. .
\end{equation}
Thus for the BB84 protocol, one has 
\begin{equation}
\label{eq}I_{opt}^R=\left\{ 
\begin{array}{c}
\log _2\left( 2-\left[ 
\frac{1+\left( 1-2\csc ^22\alpha \right) E}{1-E}\right] ^2\right)
,\;\;\alpha \leq \pi /8 \\ \log _2\left( 2-\left[ \frac{1+\left( 1-2\sec
^22\alpha \right) E}{1-E}\right] ^2\right) ,\;\;\alpha \geq \pi /8
\end{array}
\right. .
\end{equation}
For $\alpha =\pi /8$, Eq.\thinspace (105) produces Fig.\thinspace 6 of [11], as it must. Also, $I_{opt}^R$
in Eq.\thinspace (105) increases as $\alpha $ decreases below $\pi /8$, or increases above $\pi /8$. As is to be
expected, it is also evident from Eq. (105). that the standard BB84 protocol with $\alpha =\pi /8$ yields less
information than for any other value of $\alpha$. 

\section{OPTIMIZATION COMPARISON}

As reviewed above in Section 2, Slutsky, et al [11] had earlier argued that for the standard BB84 protocol
(with $\alpha =\pi /8$), the optimum set of probe parameters is given by (See Eqs.\thinspace (16) of
Ref.\thinspace [11].):  
\begin{equation}
\label{eq}\left\{ \lambda ,\mu ,\theta ,\phi ;\;\lambda =0,\mu =0,\cos
2\theta =1,\sin 2\phi =1-4E\right\} .
\end{equation}
In obtaining Eq.\thinspace (106), Slutsky et al made certain simplifying assumptions, based on the
algebraic form of the overlap function, which yielded the correct maximum Renyi information gain, but an
incomplete set of optimum probe parameters. In this section, a detailed comparison is made between the
optimization of Ref.\thinspace [11] and the complete optimization of Section 5. 

A solution to Eqs.\thinspace (33)--(35), and (28) is 
\begin{equation}
\label{eq}\left\{ \lambda ,\mu ,\theta ,\phi ;\sin \lambda =0,\cos 2\theta
=1,\sin 2\phi =1-4E\right\} .
\end{equation}
Note that Eqs.\thinspace (107) and (10) give $c=0$, consistent with the above. Since $\mu $ enters
Eqs.\thinspace (26) and (24) only through the term $\sin ^2\lambda \sin 2\mu \;$in Eqs.\thinspace (8) and
(9), and since $\sin 2\mu \sin ^2\lambda =0$, the choice sin2$\mu =0$ yields a possible solution, which
when combined with Eq.\thinspace (107) gives the set  
\begin{equation}
\label{eq}\left\{ \lambda ,\mu ,\theta ,\phi ;\;\sin 2\mu =0,\sin \lambda
=0,\cos 2\theta =1,\sin 2\phi =1-4E\right\} ,
\end{equation}
consistent with Eq.\thinspace (106) and a subset of Eq.\thinspace (90).

It is to be noted that a more general solution to Eqs.\thinspace (33)--(35) is given by 
\begin{equation}
\label{eq}\left\{ \lambda ,\mu ,\theta ;\cos 2\theta =1\right\} ,
\end{equation}
which when combined with Eq.\thinspace (28) yields 
\begin{equation}
\label{eq}\left\{ \lambda ,\mu ,\theta ,\phi ;\;\cos 2\theta =1,\;\sin 2\mu
\sin ^2\lambda =1-4E-\cos ^2\lambda \sin 2\phi \;\right\} ,
\end{equation}
coinciding with Eq.\thinspace (90). One therefore sees that even with the assumptions of Ref.\thinspace
[11], a more general set than Eq.\thinspace (106) obtains, namely, Eq.\thinspace (110). Furthermore, to
obtain a more complete optimization, one must consider the case  
\begin{equation}
\label{eq}\cos \lambda =0,
\end{equation}
in which case Eq.\thinspace (28) is not defined. Instead, using Eqs.(8),
\thinspace (11) and (24), one obtains 
\begin{equation}
\label{eq}\sin ^2\lambda \sin 2\mu =1-4E+\cos ^2\lambda \left[ 1-\cos
2\theta \left( 1+\sin 2\phi \right) \right] ,
\end{equation}
and substituting Eq.\thinspace (112) in Eq.\thinspace (9), one gets 
\begin{equation}
\label{eq}b=2-4E-\cos ^2\lambda \cos 2\theta \left( 1+\sin 2\phi \right)
-\sin ^2\lambda +\cos ^2\lambda \sin 2\phi .
\end{equation}
For minimum $b$, one then requires 
\begin{equation}
\label{eq}\frac{\partial b}{\partial \phi }=0,\;\;\frac{\partial b}{\partial
\lambda }=0,\;\;\frac{\partial b}{\partial \theta }=0.
\end{equation}
Therefore substituting Eq.\thinspace (113) in Eqs.\thinspace (114), one obtains 
\begin{equation}
\label{eq}\cos ^2\lambda \left( 1-\cos 2\theta \right) \cos 2\phi =0,
\end{equation}
\begin{equation}
\label{eq}\sin \lambda \cos \lambda \left( 1+\sin 2\phi \right) \left(
1-\cos 2\theta \right) =0,
\end{equation}
\begin{equation}
\label{eq}\cos ^2\lambda \left( 1+\sin 2\phi \right) \sin 2\theta =0.
\end{equation}
One observes that Eqs.\thinspace (115)--(117) are in fact satisfied by Eq.\thinspace (111), and from
Eqs.\thinspace (111) and (112), one obtains the optimization, Eq.\thinspace (89),  
\begin{equation}
\label{eq}\left\{ \lambda ,\mu ,\theta ,\phi ;\;\cos \lambda =0,\;\sin 2\mu
=1-4E\right\} ,
\end{equation}
which is the missing set in the optimization of Ref.\thinspace [11].

It is evident that Eqs. (115)--(117) are also satisfied by $\sin 2\phi = -1$, and combining this with Eq. (112), one
obtains an additional set of optimum probe parameters:
\begin{equation}
\left\{ {\lambda ,\mu ,\theta ,\phi ;\sin 2\phi =-1,\sin 2\mu \sin ^2\lambda =1-4E+\cos ^2\lambda } \right\}\ .
\end{equation}
This solution was also not obtained in Ref. [11]. It is at this point important to note that, since the analysis in Sections 3
and 4 and Ref. [12] was performed for arbitrary $\alpha$, the possible solution given by Eq. (A-48) was ignored
because it followed from Eq. (A-173) that $\alpha = \pi /8$ and $e_\phi = -1$ are required. However, if $\alpha =
\pi /8$, then Eqs. (A-48) and (57) are satisfied for the set of probe parameters given by Eq. (119).

Also, Appendix E of Ref.\thinspace [11] addresses an alternative simplification of the optimization
problem, which is reviewed in the above at the end of Section 2. The problem is inverted so that $E$ is
minimized subject to constant $b,\arg $uing that the conditional minimum of $E$ is a monotonically
decreasing function of $b$ for the domain of interest ($0\leq $ $E<1/2$). In Eq.\thinspace (59) $E$ is
seen to be a monotonically decreasing function of $b$, as claimed in Appendix E of Ref.\thinspace [11].
Furthermore, since in this case cos2$\theta =1$, then, together with Eq.\thinspace (112), one obtains the
set of optimum probe parameters,  
\begin{equation}
\label{eq}\left\{ \lambda ,\mu ,\theta ,\phi ;\;\cos 2\theta =1,\;\sin 2\mu
\sin ^2\lambda =1-4E-\cos ^2\lambda \sin 2\phi \;\right\} ,
\end{equation}
in agreement with Eq.\thinspace (90). However, the optimization given in Ref.\thinspace [11], namely
Eq.\thinspace (106) above, is a subset of Eq.\thinspace (120). Furthermore, if the multiplier $\cos
\lambda $ of $\cos 2\theta $ in Eq.\thinspace (27) is vanishing, then $\cos 2\theta =1$ does not
necessarily produce the best $E_0$. Thus if  
\begin{equation}
\label{eq}\cos \lambda =0,
\end{equation}
then combining this with Eq.\thinspace (27) yields the optimum set, 
\begin{equation}
\label{eq}\left\{ \lambda ,\mu ,\theta ,\phi ;\;\cos \lambda =0,\;\sin 2\mu
=1-4E\right\} ,
\end{equation}
which is again the missing set, Eq.\thinspace (89).

Also, if $\sin 2\phi = -1$, then the multiplier of $\cos 2\theta$ in Eq. (27) is again vanishing, and then $\cos 2\theta = 1$
does not necessarily produce the best $E_0$, but $\sin 2\phi = -1$ along with Eq. (27) again leads to the optimization given
by Eq. (119). It can also be shown that $\cos 2\theta = 0$, for which Eq. (28) is also not satisfied, leads to no
additional optimimum sets of probe parameters.

\section{SECRECY CAPACITY}

The maximum Renyi information gain, Eq. (105), can be used to calculate the secrecy capacity of the four-state
protocol in the presence of the individual attack. Before calculating the secrecy capacity, a review
of key distillation is at this point appropriate. Let
$m$ bits of raw data be received by the legitimate receiver in the four-state quantum-key-distribution protocol,
and suppose $n$ bits of sifted data remain following removal of $(m-n)$ inconclusive bits, and suppose there are
$e_T$ bits of erroneous data, leaving $(n-e_T)$ bits of corrected data. Corrected data includes data remaining after
discarding inconclusive results and also erroneous data as determined by block checksums and bisective
search. Privacy amplification is the procedure for obtaining a more secure, but shorter, key. This is
achieved by removing from the $(n-e_T)$ bits of corrected data a number $s$ of bits (the privacy
amplification compression level) that is the sum of the possible contributions to information leakage.
There then remain $(n-e_T-s)$ bits, and this is the size of the final key. The privacy amplification
compression level $s$ is given by [19]  
\begin{equation}
\label{eq}s=t(n,e_T)+q+\nu +g,
\end{equation}
where $q$ is the estimated information leakage during error correction, $\nu $ is the estimated leakage
from any multi-photon bits, $g$ is an extra safety margin, and $t(n,e_T)$ is the defense function. The
defense function, in general, depends on the size $n$ of the sifted data, and on the number $e_T$ of
errors, and is chosen appropriately by the legitimate users, in order to effectively defend against an
eavesdropping attack. The defense function $t(n,e_T)$ is the estimated upper bound on possible information
leakage through eavesdropping on the quantum channel. Quantitatively it is determined by the maximum total
Renyi information gain $I_T^{R\text{ }}$by the eavesdropping probe. (It is proved in [11] that the optimum
individual attack maximizes both the Renyi and Shannon information gain by the eavesdropping probe.) The
maximum Renyi information gain by the eavesdropper is based on minimizing the overlap of the measured
probe states correlated with the disturbed signal states of the legitimate receiver, conditional on fixed
induced error rate. The compression level $s$ must be chosen so that the probability is small that
$I_T^R>t(n,e_T)$. An attack is successful if it introduces $e_T$ errors on the $n$ bits of sifted data,
and yields a Renyi information $I_T^R>t(n,e_T)$ on the $(n-e_T)$ bits of corrected data. The probability
of a successful attack must be negligible. In the presence of noise and channel losses, it is not
sufficient, for the security of a quantum key distribution system, to detect eavesdropping. It must be
insured that the shared data is sufficiently secure. 

It is well to recall the privacy amplification theorem [9]. First, however, recall the definition of the
Renyi information $I^R(l)$ on an $l$ bit string $X$ having probability distribution $P_X(X)$,
namely,\thinspace  
\begin{equation}
\label{eq}I^R(l)=l+\log _2\left\langle P_X(X)\right\rangle =l+\log _2
\stackunder{X}{\sum }P_X^2(X), 
\end{equation}
where the bracket denotes the expectation value. ($P_X^2(X)$ is often referred to as the collision
probability.) The privacy amplification theorem states that if the eavesdropper's Renyi information gain
$I^R(l)$ on an $l$ bit data string is less than some quantity $r$, namely,  
\begin{equation}
\label{eq}I^R(l)\leq r, 
\end{equation}
then the eavesdropper's Shannon information $I^H(l-s)$ on the reduced $(l-s)$ bit string, averaged over
the choice of privacy amplification hash function, is bounded above, namely,  
\begin{equation}
\label{eq}\left\langle I^H(l-s)\right\rangle \leq \frac 1{\ln 2}2^{r-s}, 
\end{equation}
where here the brackets denote the average. By choosing the compression level $s$ sufficiently large, the
exponent on the right hand side of Eq.\thinspace (126) becomes sufficiently negative that the average
Shannon information can be made arbitrarily small. Thus, given an upper bound on the Eavesdropper's Renyi
information gain, the corrected data can be subjected to the reduction procedure of privacy amplification
to yield an even shorter string on which the eavesdropper's Shannon information is arbitrarily low. The
secrecy of the final key is recovered (but reduced in size) if an upper bound can be determined on the
maximum Renyi information gain by the eavesdropper on corrected data. 

The average secrecy capacity $C_s^{\prime }$ of a quantum cryptosystem is the number of secret bits
produced per bit from the transmitter, and is given by 

\begin{equation}
\label{eq}C_s^{\prime }=\stackunder{m\rightarrow \infty }{Lim}\left\langle 
\frac{n-e_T-s}m\right\rangle . 
\end{equation}
Here the limit of a very long transmission is understood in which $m$, the number of bits of raw data, is
very large. 

The numerator of Eq.\thinspace (127), $\left( n-e_T-s\right) ,$ is the size of the final key, where $n$ is
the number of bits of sifted data with the inconclusive bits removed, $e_T$ is the number of bits of
erroneous discarded data due to error correction, and $s$ is the privacy amplification compression level.
The average secrecy capacity, Eq.\thinspace (127),  converges in distribution to [19] 

\begin{equation}
\label{eq}C_s^{\prime }=\left\langle \frac nm\right\rangle \left(
1-\left\langle \frac{e_T}n\right\rangle -\frac{t_F}n|_{\frac{e_T}
n=\left\langle \frac{e_T}n\right\rangle }-\stackunder{m\rightarrow \infty }{
Lim}\left\langle \frac qn\right\rangle \right) .
\end{equation}
The factor $\left\langle \frac nm\right\rangle $ in Eq.\thinspace (128) is the conclusive rate. Since the
inconclusive rate $R_{?}$ is 1/2 for the BB84 protocol [11,19], and remains unchanged in the presence of
the individual attack, the conclusive rate must also be 1/2, namely,  
\begin{equation}
\label{eq}\left\langle \frac nm\right\rangle =(1-R_{?})=\left( 1-\frac
12\right) =\frac 12.
\end{equation}
Also in Eq.\thinspace (128), $\left\langle \frac{e_T}n\right\rangle $ is the average intrinsic error rate,
and $\left\langle \frac qn\right\rangle $ is the average information leakage during error correction.
Since the present work focuses on the information leakage through eavesdropping (represented by the third
term in Eq.\thinspace (128)), possible additional terms, $\left( -\left\langle \nu /n\right\rangle \right)
$ and $\left( -\left\langle g/n\right\rangle \right) $, are dropped in Eq.\thinspace (128) (See
Eq.\thinspace (123)). In the third term of Eq.\thinspace (128), $\frac{t_F}n|_{\frac{e_T}n=\left\langle
\frac{e_T}n\right\rangle }$ is the average defense frontier $t_F$ evaluated at the average intrinsic error
rate. In the individual attack, each signal is attacked individually and in the same way, and it is
assumed that the signal states, errors, and measurement outcomes of the probe and the legitimate receiver
are all independently and identically distributed [19]. Multiple eavesdropping strategies are considered
with different induced error rates, but the attack is restricted to the set of strategies yielding the
greatest attainable expected Renyi information gain for a given expected error rate. The defense frontier
$t_F$ is, for all possible eavesdropping strategies, the upper bound on the information leakage through
eavesdropping, based on an optimal eavesdropper in the limit of a long transmission. The defense frontier
$t_F$ is chosen to minimize the chance of any successful eavesdropping strategy, and, for the individual
attack, it is given by [19] 

\begin{equation}
\label{eq}t_F(n,e_T)=\stackunder{e\leq e_T}{\;\max }\left\{ n\left( 1-\frac
en\right) I_{opt}^R\left( \frac en+\xi \right) +\xi \left[ n^2\left( 1-\frac
en\right) \right] ^{1/2}\right\} ,
\end{equation}
where $I_{opt}^R\left( E\right) $ is the maximum Renyi information gain on corrected data by the
eavesdropping probe, and conditional on fixed error rate $E=(e/n)$; and $\xi $ is defined by  
\begin{equation}
\label{eq}\xi =\frac 1{\left( 2n\right) ^{1/2}}\limfunc{erf}{}^{-1}(1-p),
\end{equation}
where $\limfunc{erf}^{-1}$ denotes the inverse standard error function. The standard error function
$\limfunc{erf}(z)$ is defined by  
\begin{equation}
\label{eq}\limfunc{erf}(z)=\frac 2{\sqrt{\pi }}\int\limits_0^ze^{-y^2}dy.
\end{equation}

Also in Eq.\thinspace (131), $p$ is the probability for successful eavesdropping ($I_R^T>t(n,e_T)$) on
($n-e_T)$ bits of corrected data and producing $e_T$ errors; and $p$ can be made arbitrarily small. The
defense frontier, Eq.\thinspace (130), was determined by Slutsky, Rao, Sun, Tancevski, and Fainman [19] by
clever use of the central limit theorem of probability theory, and is constructed to minimize the chance
of successful eavesdropping. Using Eqs.\thinspace (128)--(130), the asymptotic secrecy capacity, in the
limit of long transmission with $m\rightarrow \infty $, $n\rightarrow \infty $, and $\xi \rightarrow 0$,
and for $q=0$, becomes [19] 

\begin{equation}
\label{eq}C_s^{\prime }|_{q=0,n\rightarrow \infty ,\xi \rightarrow 0}=\frac
12\left( 1-E-\stackunder{E^{^{\prime }}\leq E}{\;\max }(1-E^{\prime
})I_{opt}^R(E^{\prime })\right) ,
\end{equation}
where $E$ is the error rate, and ($\stackunder{x^{\prime }\leq x}{\max \,}f(x^{\prime })$) denotes the
maximum value of a function $f(x^{\prime })$ for $x^{\prime }\leq x$. Also in Eq.\thinspace (133),
$I_{opt}^R(E^{\prime })$ is the maximum Renyi information gain on corrected data by the eavesdropping
probe, conditional on fixed error rate $E^{\prime }$. The asymptotic secrecy capacity, Eq.\thinspace
(133), is based on the definition of average secrecy capacity, Eq.\thinspace (127), as given in the
literature [19], however it is important to emphasize that the condition of maximum Renyi information gain
by the eavesdropper may be overly conservative (See Section VI of Bennett, et al [9]). 

Substituting Eq.\thinspace (105) in Eq.\thinspace (133), one obtains for the asymptotic secrecy capacity
[13]: 

\begin{equation}
\label{eq}
\begin{array}{c}
C_s^{\prime }|_{q=0,n\rightarrow \infty ,\xi \rightarrow
0}=\;\;\;\;\;\;\;\;\;\;\;\;\;\;\;\;\;\;\;\;\;\;\;\;\;\;\;\;\;\;\;\;\;\;\;\;
\;\;\;\;\;\;\;\;\;\;\;\;\;\;\;\;\;\;\;\;\;\;\;\;\;\;\;\;\;\;\;\;\;\; \\ 
\;\;\;\left\{ 
\begin{array}{c}
\frac 12\left( 1-E-
\stackunder{E^{^{\prime }}\leq E}{\;\max }(1-E^{\prime })\log _2\left[
2-\left( \frac{1+\left( 1-2\csc ^22\alpha \right) E^{^{\prime }}}{
1-E^{^{\prime }}}\right) ^2\right] \right) ,\;\;\;\alpha \leq \pi /8 \\ 
\frac 12\left( 1-E-\stackunder{E^{^{\prime }}\leq E}{\;\max }(1-E^{\prime
})\log _2\left[ 2-\left( \frac{1+\left( 1-2\sec ^22\alpha \right)
E^{^{\prime }}}{1-E^{^{\prime }}}\right) ^2\right] \right) ,\;\;\;\alpha
\geq \pi /8
\end{array}
\right. .
\end{array}
\end{equation}
For $\alpha =\pi /8$, Eq.\thinspace (134) also agrees with [19]. It is evident from Eqs.\thinspace (104),
(105), and (134) that as a function of $\alpha $, for fixed error rate, the overlap of correlated probe
states is greatest, the Renyi information gain by the probe is least, and the secrecy capacity is greatest
for $\alpha =\pi /8$, which corresponds to the standard BB84 protocol [3] with $\stackrel{\_}{\theta }=\pi
/4$. 

\section{SUMMARY}

The maximum Renyi information gain, Eq.\thinspace (105), by a Fuchs-Peres probe [11,15] is calculated for
varying angle between the signal bases in the four-state protocol [3] of quantum key distribution. Two
sets of optimized probe parameters, Eqs.\thinspace (99) and (100) for $\alpha \leq \pi /8$, and
Eqs.\thinspace (101) and (102) for $\alpha \geq \pi /8$, are found to yield the optimization. Only a
subset of one of these sets was found previously [11], for $\alpha =\pi /8$ (Eq.\thinspace (100) with
$\sin \lambda =0$ and $\alpha =\pi /8$, or equivalently Eq.\thinspace (90) with $\sin \lambda =0$). When
the angle between the signal bases is the standard 45 degrees ($\alpha =\pi /8$), the result of Slutsky,
Rao, Sun, and Fainman [11] is recovered. Also, it was shown by explicit calculations that Eq.\thinspace
(105) gives the maximum information gain by the probe for a representative range of values of $\alpha $.
Also, the maximum Renyi information, Eq.\thinspace (105), for constant error rate, increases as $\alpha $
decreases below $\pi /8$, or increases above $\pi /8$. 

Also for $\alpha = \pi/8$, an additional set of optimum probe parameters, Eq. (119) previously ignored, has been
found. A detailed comparison has been made between the optimizations of Ref.\thinspace [12] and Ref.\thinspace
[11]. The reasons why the optimization of Ref.\thinspace [11] (Eq.\thinspace (106) above) did not yield the
complete set of optimum probe parameters are because, in one approach considered there, the restriction in
applicability of Eq.\thinspace (28) was ignored. Also the parameter $\mu$ was unecessively restricted in Eq. (106). 

Next, following a review of the process of key distillation from the quantum transmission in quantum key
distribution, the asymptotic secrecy capacity, Eq.\thinspace (134), of the four-state protocol has been
calculated for the case of an individual attack in which the eavesdropping probe is entangled with the
signal states, and states of the probe become correlated with the states measured by the legitimate
receiver. The calculation generalizes earlier work to include an arbitrary angle between the signal bases.\\ 
ACKNOWLEDGEMENTS

This work was supported by the U.S. Army Research Laboratory and the Defense Advanced Research Projects
Agency. Useful communications with J. Myers, S. J. Lomonaco, J. D. Franson, J. D. Murley, and M. Foster
are gratefully acknowledged. The author wishes to especially thank Prof. Young S. Kim for inviting him to
organize the sessions on quantum computing and quantum communication in which this paper was presented at the {\it First
Feynman Festival}, 23--28 August, 2002 at the University of Maryland, College Park, MD. 

\cleardoublepage
\renewcommand\theequation{\thesection-\arabic{equation}}
\appendix\section{Possible Extrema}
\setcounter{equation}{0}

In this appendix, the sets of conditions for the existence of possible extrema of the overlap of correlated probe
states are determined by using Eqs. (69)--(71). Also, the possible extrema and the associated probe parameters are
calculated. First, substituting Eq.\thinspace (68) in Eq.\thinspace (69), one obtains 
\begin{equation}
\label{eq}\frac{\partial q}{\partial \lambda }+\frac{q-1+2E}{\left[ 4\left(
1-E\right) ^2-c^2\sin ^22\alpha \right] }\sin ^22\alpha \,c\frac{\partial c}{
\partial \lambda }=0. 
\end{equation}
Using Eqs.\thinspace (67) and (10), it follows that 
\begin{equation}
\label{eq}
\begin{array}{c}
\frac{\partial q}{\partial \lambda }=-2\left( \cos \lambda \sin \lambda
\right) \left\{ \left( 2-\tan ^22\alpha \right) \left[ \cot ^22\alpha -\cos
2\theta \left( \sin 2\phi +\cot ^22\alpha \right) \right] \right. \\ +\left.
\sin 2\phi \left[ 1+\left( 1-\tan ^22\alpha \right) \cos 2\theta \right]
\right\} , 
\end{array}
\end{equation}
\begin{equation}
\label{eq}c\frac{\partial c}{\partial \lambda }=-2\cos ^3\lambda \sin
\lambda \sin ^22\theta \cos ^22\phi . 
\end{equation}

Then substituting Eqs.\thinspace (A-2) and (A-3) in Eq.\thinspace (A-1), one requires 
\begin{equation}
\label{eq}\sin \lambda \cos \lambda F_1(\lambda ,\theta ,\phi )=0, 
\end{equation}
where 
\begin{equation}
\label{eq}
\begin{array}{c}
F_1(\lambda ,\theta ,\phi )=2\left\{ \left( 2-\tan ^22\alpha \right) \left[
\cot ^22\alpha -\cos 2\theta \left( \sin 2\phi +\cot ^22\alpha \right)
\right] \right. \\ 
+\left. \sin 2\phi \left[ 1+\left( 1-\tan ^22\alpha \right) \cos 2\theta
\right] \right\} \\ 
\;\;\;\;\;\;\;\;\;\;\;\;\;\;\;\;\;\;\;\;\;+\frac{2\left( q-1+2E\right) }{
4(1-E)^2-c^2\sin ^22\alpha }\sin ^22\alpha \cos ^2\lambda \sin ^22\theta
\cos ^22\phi . 
\end{array}
\end{equation}

Next, substituting Eq.\thinspace (68) in Eq.\thinspace (70), one obtains 
\begin{equation}
\label{eq}\frac{\partial q}{\partial \theta }+\frac{q-1+2E}{\left[ 4\left(
1-E\right) ^2-c^2\sin ^22\alpha \right] }\sin ^22\alpha \,c\frac{\partial c}{
\partial \theta }=0. 
\end{equation}
Using Eqs.\thinspace (67) and (10), it follows that 
\begin{equation}
\label{eq}\frac{\partial q}{\partial \theta }=2\sin 2\theta \cos ^2\lambda
\left( \sin 2\phi +2\cot ^22\alpha -1\right) , 
\end{equation}
\begin{equation}
\label{eq}c\frac{\partial c}{\partial \theta }=2\sin 2\theta \cos ^4\lambda
\cos 2\theta \cos ^22\phi . 
\end{equation}
Then substituting Eqs.\thinspace (A-7) and (A-8) in Eq.\thinspace (A-6), one requires 
\begin{equation}
\label{eq}\sin 2\theta \cos ^2\lambda F_2(\lambda ,\theta ,\phi )=0, 
\end{equation}
where 
\begin{equation}
\label{eq}
\begin{array}{c}
F_2(\lambda ,\theta ,\phi )=2\left( \sin 2\phi +2\cot ^22\alpha -1\right)
\;\;\;\;\;\;\;\;\;\;\;\;\;\;\;\;\;\;\;\;\;\;\;\;\;\;\;\;\;\;\;\;\;\;\;\;\;\;
\;\;\;\;\;\;\;\; \\ 
+\;\frac{2\left( q-1+2E\right) }{\left[ 4(1-E)^2-c^2\sin ^22\alpha \right] }
\sin ^22\alpha \cos ^2\lambda \cos 2\theta \cos ^22\phi . 
\end{array}
. 
\end{equation}

Next, substituting Eq.\thinspace (68) in Eq.\thinspace (71), one obtains 
\begin{equation}
\label{eq}\frac{\partial q}{\partial \phi }+\frac{q-1+2E}{\left[ 4\left(
1-E\right) ^2-c^2\sin ^22\alpha \right] }\sin ^22\alpha \,c\frac{\partial c}{
\partial \phi }=0. 
\end{equation}
Using Eqs.\thinspace (67) and (10), one gets 
\begin{equation}
\label{eq}\frac{\partial q}{\partial \phi }=2\cos ^2\lambda \cos 2\phi
\left( 1-\cos 2\theta \right) , 
\end{equation}
\begin{equation}
\label{eq}c\frac{\partial c}{\partial \phi }=-2\cos ^4\lambda \sin ^22\theta
\sin 2\phi \cos 2\phi . 
\end{equation}
Then substituting Eqs.\thinspace (A-12) and (A-13) in Eq.\thinspace (A-11), one requires 
\begin{equation}
\label{eq}\cos ^2\lambda \cos 2\phi F_3(\lambda ,\theta ,\phi )=0, 
\end{equation}
where 
\begin{equation}
\label{eq}
\begin{array}{c}
F_3(\lambda ,\theta ,\phi )=2\left( 1-\cos 2\theta \right)
\;\;\;\;\;\;\;\;\;\;\;\;\;\;\;\;\;\;\;\;\;\;\;\;\;\;\;\;\;\;\;\;\;\;\;\;\;\;
\;\;\; \\ 
-\;\frac{2\left( q-1+2E\right) }{\left[ 4(1-E)^2-c^2\sin ^22\alpha \right] }
\sin ^22\alpha \cos ^2\lambda \sin ^22\theta \sin 2\phi . 
\end{array}
\end{equation}

Summarizing Eqs.\thinspace (A-4), (A-9), and (A-14), possible extrema of the overlap of correlated probe
states are determined by  
\begin{equation}
\label{eq}\text{(a)}\;\;\;\;\;\;\;\;\;\;\;\sin \lambda \cos \lambda
F_1(\lambda ,\theta ,\phi )=0, 
\end{equation}
\begin{equation}
\label{eq}\text{(b)}\;\;\;\;\;\;\;\;\;\sin 2\theta \cos ^2\lambda
F_2(\lambda ,\theta ,\phi )=0, 
\end{equation}
\begin{equation}
\label{eq}\text{(c)\ \ \ \ \ \ \ \ }\cos ^2\lambda \cos 2\phi F_3(\lambda
,\theta ,\phi )=0. 
\end{equation}
Three possible ways of satisfying Eq.\thinspace (A-16) are 
\begin{equation}
\label{eq}\text{(a1)}\;\;\;\;\;\;\;\;\;\;\;\sin \lambda =0, 
\end{equation}
\begin{equation}
\label{eq}\text{(a2)}\;\;\;\;\;\;\;\;\;\;\;\cos \lambda =0, 
\end{equation}
\begin{equation}
\label{eq}\text{(a3)\ \ \ \ \ \ \ \ \ \ \ \ \ \ }F_1=0. 
\end{equation}

Two possible ways of satisfying Eq.\thinspace (A-19) and (A-17) are 
\begin{equation}
\label{eq}\text{(a11)}\;\;\;\;\;\;\;\;\;\;\;\sin \lambda =0,\;\;\sin 2\theta
=0, 
\end{equation}
\begin{equation}
\label{eq}\text{(a12)}\;\;\;\;\;\;\;\;\;\;\;\sin \lambda =0,\;\;\;\;\;F_2=0. 
\end{equation}
Two possible ways of satisfying Eqs.\thinspace (A-22) and (A-18), and therefore also Eqs.\thinspace (A-16) and
(A-17), are  
\begin{equation}
\label{eq}\text{(a111)}\;\;\;\;\;\;\;\;\;\;\;\sin \lambda =0,\;\;\;\;\sin
2\theta =0,\;\;\cos 2\phi =0, 
\end{equation}
\begin{equation}
\label{eq}\text{(a112)}\;\;\;\;\;\;\;\;\;\;\;\sin \lambda =0,\;\;\;\;\sin
2\theta =0,\;\;\;\;\;\;\;F_3=0. 
\end{equation}

Two possible ways of satisfying Eqs.\thinspace (A-23) and (A-18), and therefore also Eqs.\thinspace (A-16) and
(A-17), are  
\begin{equation}
\label{eq}\text{(a121)}\;\;\;\;\;\;\;\;\;\;\;\sin \lambda =0,\;\;\;\;\cos
2\phi =0,\;\;\;\;\;\;\;F_2=0, 
\end{equation}
\begin{equation}
\label{eq}\text{(a122)}\;\;\;\;\;\;\;\;\;\;\;\sin \lambda
=0,\;\;\;\;\;\;\;\;\;F_2=0,\;\;\;\;\;\;\;F_3=0. 
\end{equation}

Equation (A-20) satisfies Eq.\thinspace (A-17) and (A-18). Therefore, another way of satisfying Eqs.\thinspace
(A-16)--(A-18) is  
\begin{equation}
\label{eq}\text{(a2)}\;\;\;\;\;\;\;\;\;\;\;\cos \lambda =0. 
\end{equation}

Three possible ways of satisfying Eqs.\thinspace (A-21) and (A-17) are 
\begin{equation}
\label{eq}\text{(c1)}\;\;\;\;\;\;\;\;\;\;\;F_1=0,\;\;\;\;\;\sin 2\theta =0, 
\end{equation}
\begin{equation}
\label{eq}\text{(c2)}\;\;\;\;\;\;\;\;\;\;\;\;F_1=0,\;\;\;\;\;\cos \lambda
=0, 
\end{equation}
\begin{equation}
\label{eq}\text{(c3)}\;\;\;\;\;\;\;\;\;\;\;\;\;F_1=0,\;\;\;\;\;\;\;\;F_2=0. 
\end{equation}
Three possible ways of satisfying Eqs.\thinspace (A-29) and (A-18), and therefore also Eqs.\thinspace (A-16)
and (A-17), are  
\begin{equation}
\label{eq}\text{(c11)}\;\;\;\;\;\;\;\;\;\;\;F_1=0,\;\;\;\;\;\;\sin 2\theta
=0,\;\;\;\;\;\;\cos 2\phi =0, 
\end{equation}
\begin{equation}
\label{eq}\text{(c12)}\;\;\;\;\;\;\;\;\;\;\;\;F_1=0,\;\;\;\;\;\sin 2\theta
=0,\;\;\;\;\;\;\cos \lambda =0, 
\end{equation}
\begin{equation}
\label{eq}\text{(c13)}\;\;\;\;\;\;\;\;\;\;\;F_1=0,\;\;\;\;\;\;\sin 2\theta
=0,\;\;\;\;\;\;\;\;\;\;F_3=0. 
\end{equation}
Eq.\thinspace (A-30) satisfies Eq.\thinspace (A-18), and therefore, another way of satisfying Eqs.\thinspace
(A-16)--(A-18) is  
\begin{equation}
\label{eq}\text{(c2)}\;\;\;\;\;\;\;\;\;\;\;F_1=0,\;\;\;\;\;\cos \lambda =0. 
\end{equation}
Three possible ways of satisfying Eqs.\thinspace (A-31) and (A-18), and therefore also Eqs.\thinspace (A-16)
and (A-17), are  
\begin{equation}
\label{eq}\text{(c31)}\;\;\;\;\;\;\;\;\;\;\;F_1=0,\;\;\;\;\;\;\;\;\;\;
\;F_2=0,\;\;\;\;\;\;\cos 2\phi =0, 
\end{equation}
\begin{equation}
\label{eq}\text{(c32)}\;\;\;\;\;\;\;\;\;\;\;\;F_1=0,\;\;\;\;\;\;\;\;\;\;
\;F_2=0,\;\;\;\;\;\;\cos \lambda =0, 
\end{equation}
\begin{equation}
\label{eq}\text{(c33)}\;\;\;\;\;\;\;\;\;\;\;\;F_1=0,\;\;\;\;\;\;\;\;\;\;\;\;
\;F_2=0,\;\;\;\;\;\;\;\;\;F_3=0. 
\end{equation}

Summarizing Eqs.\thinspace (A-24)--(A-28) and (A-32)--(A-38), possible solutions to Eqs.\thinspace (A-16)--(A-18)
are determined by  
\begin{equation}
\label{eq}(\text{A})\;\;\;\;\sin \lambda =0,\;\;\;\sin 2\theta =0,\;\;\;\cos
2\phi =0, 
\end{equation}
\begin{equation}
\label{eq}(\text{B})\;\;\;\;\sin \lambda =0,\;\;\;\sin 2\theta
=0,\;\;\;\;\;\;\;F_3=0, 
\end{equation}
\begin{equation}
\label{eq}(\text{C})\;\;\;\;\sin \lambda =0,\;\;\;\cos 2\phi
=0,\;\;\;\;\;\;\;F_2=0, 
\end{equation}
\begin{equation}
\label{eq}(\text{D})\;\;\;\;\sin \lambda
=0,\;\;\;\;\;\;\;\;\;F_2=0,\;\;\;\;F_3=0,\; 
\end{equation}
\begin{equation}
\label{eq}(\text{E})\;\;\;\;\cos \lambda
=0,\;\;\;\;\;\;\;\;\;\;\;\;\;\;\;\;\;\;\;\;\;\;\;\;\;\;\;\;\;\;\;\;\;\;\;\;
\;\; 
\end{equation}
\begin{equation}
\label{eq}(\text{F})\;\;\;\sin 2\theta =0,\;\;\;\;\;\cos 2\phi
=0,\;\;\;\;F_1=0, 
\end{equation}
\begin{equation}
\label{eq}(\text{G})\;\;\;\cos \lambda =0,\;\;\;\;\sin 2\theta
=0,\;\;\;\;F_1=0,\; 
\end{equation}
\begin{equation}
\label{eq}(\text{H})\;\;\;\sin 2\theta
=0,\;\;\;\;\;\;\;\;\;F_1=0,\;\;\;\;F_3=0, 
\end{equation}
\begin{equation}
\label{eq}(\text{I})\;\;\;\;\cos \lambda
=0,\;\;\;\;\;\;\;\;\;F_1=0,\;\;\;\;\;\;\;\;\;\;\;\;\;\;\; 
\end{equation}
\begin{equation}
\label{eq}(\text{J})\;\;\;\cos 2\phi
=0,\;\;\;\;\;\;\;\;\;F_1=0,\;\;\;\;F_2=0, 
\end{equation}
\begin{equation}
\label{eq}(\text{K})\;\;\;\cos \lambda
=0,\;\;\;\;\;\;\;\;\;F_1=0,\;\;\;\;F_2=0, 
\end{equation}
\begin{equation}
\label{eq}(\text{L})\;\;\;\;\;\;\;F_1=0,\;\;\;\;\;\;\;\;\;F_2=0,\;\;\;
\;F_3=0. 
\end{equation}

First consider possible extrema determined by possibility (B), Eq.\thinspace (A-40): 
\begin{equation}
\label{eq}\sin \lambda =0,
\end{equation}
\begin{equation}
\label{eq}\sin 2\theta =0,
\end{equation}
\begin{equation}
\label{eq}F_3=0.
\end{equation}
From Eqs.\thinspace (A-15), (A-52) and (A-53), it follows that 
\begin{equation}
\label{eq}\cos 2\theta =1.
\end{equation}
Substituting Eqs.\thinspace (A-51) and (A-54) in Eq.\thinspace (66), it follows that 
\begin{equation}
\label{eq}\sin 2\phi =1-2E\csc ^22\alpha .
\end{equation}
Next substituting Eqs.\thinspace (A-51), (A-52), and (A-54) in Eqs.\thinspace (8)--(11), one obtains 
\begin{equation}
\label{eq}a=\sin 2\phi ,
\end{equation}
\begin{equation}
\label{eq}b=\sin 2\phi ,
\end{equation}
\begin{equation}
\label{eq}c=0,
\end{equation}
\begin{equation}
\label{eq}d=1.
\end{equation}
Then substituting Eqs.\thinspace (A-55)--(A-59) in Eq.\thinspace (62), one obtains 
\begin{equation}
\label{eq}Q=\frac{1+\left( 1-2\csc ^22\alpha \right) E}{1-E}.
\end{equation}
For $\alpha =\pi /8$, Eq.\thinspace (A-60) becomes Eq.\thinspace (57), corresponding to the standard BB84
optimization [11], as must be the case. 

Next, consider possibility (A), given by Eq.\thinspace (A-39): 
\begin{equation}
\label{eq}\sin \lambda =0, 
\end{equation}
\begin{equation}
\label{eq}\sin 2\theta =0, 
\end{equation}
\begin{equation}
\label{eq}\cos 2\phi =0. 
\end{equation}
From Eqs.\thinspace (A-62) and (A-63), it follows that 
\begin{equation}
\label{eq}\cos 2\theta =e_\theta , 
\end{equation}
and 
\begin{equation}
\label{eq}\sin 2\phi =e_\phi ,. 
\end{equation}
where 
\begin{equation}
\label{eq}e_\theta =\pm 1,\;\;\;\;e_\phi =\pm 1. 
\end{equation}
Substituting Eqs.\thinspace (A-61), (A-64), and (A-65) in Eq.\thinspace (65), then one requires 
\begin{equation}
\label{eq}E=\frac 12\left[ 1-e_\theta +e_\theta \left( 1-e_\phi \right) \sin
^22\alpha \right] . 
\end{equation}
Next substituting Eqs.\thinspace (A-61)--(A-66) in Eqs.\thinspace (8)--(11), one obtains 
\begin{equation}
\label{eq}a=e_\theta e_\phi , 
\end{equation}
\begin{equation}
\label{eq}b=e_\phi , 
\end{equation}
\begin{equation}
\label{eq}c=0, 
\end{equation}
\begin{equation}
\label{eq}d=e_\theta . 
\end{equation}
Then substituting Eqs.\thinspace (A-67)--(A-71) in Eq.\thinspace (62), one obtains 
\begin{equation}
\label{eq}Q=\frac{e_\phi \left( 1+e_\theta \right) +e_\theta \left( 1-e_\phi
\right) \sin ^22\alpha }{\left( 1+e_\theta \right) -e_\theta \left( 1-e_\phi
\right) \sin ^22\alpha }. 
\end{equation}
For $e_\theta =\pm 1$ and $e_\phi =+1$, Eq.\thinspace (A-72) yields 
\begin{equation}
\label{eq}Q=1. 
\end{equation}
For $e_\theta =\pm 1$ and $e_\phi =-1$, Eq.\thinspace (A-72) yields 
\begin{equation}
\label{eq}Q=-1. 
\end{equation}
One concludes that possibility (A), Eq.\thinspace (A-39), does not yield the minimum overlap.

Next, consider possibility (C), given by Eq.\thinspace (A-41): 
\begin{equation}
\label{eq}\sin \lambda =0, 
\end{equation}
\begin{equation}
\label{eq}\cos 2\phi =0, 
\end{equation}
\begin{equation}
\label{eq}F_2=0. 
\end{equation}
Next, substituting Eqs.\thinspace (A-10) and (A-76) in Eq.\thinspace (A-77), one obtains 
\begin{equation}
\label{eq}\sin 2\phi =1-2\cot ^22\alpha . 
\end{equation}
Then combining Eqs.\thinspace (A-76) and (A-78), one requires: 
\begin{equation}
\label{eq}\cot ^22\alpha =\frac 12\left( 1-e_\phi \right) , 
\end{equation}
and therefore, using Eq.\thinspace (A-66), one requires $e_\phi =-1$, and 
\begin{equation}
\label{eq}\alpha =\pi /8. 
\end{equation}
Furthermore, using Eqs.\thinspace (A-75), (A-76), and (A-80) in Eq.\thinspace (66), one requires: 
\begin{equation}
\label{eq}E=\frac 12. 
\end{equation}
Therefore possibility (C) does not yield a solution.

Next consider possibility (D), given by Eq.\thinspace (A-42): 
\begin{equation}
\label{eq}\sin \lambda =0,
\end{equation}
\begin{equation}
\label{eq}F_2=0,
\end{equation}
\begin{equation}
\label{eq}F_3=0.
\end{equation}
Using Eqs.\thinspace (A-82) and (10), one has 
\begin{equation}
\label{eq}c=\sin 2\theta \cos 2\phi .
\end{equation}
Also, using Eqs.\thinspace (A-83) and (A-10), one requires 
\begin{equation}
\label{eq}\left[ \frac{q-1+2E}{4\left( 1-E\right) ^2-c^2\sin ^22\alpha }
\right] =\frac{1-2\cot ^22\alpha -\sin 2\phi }{\sin ^22\alpha \cos 2\theta
\cos ^22\phi }.
\end{equation}
Also, Eqs.\thinspace (A-84) and (A-15) require 
\begin{equation}
\label{eq}\left[ \frac{q-1+2E}{4\left( 1-E\right) ^2-c^2\sin ^22\alpha }
\right] =\frac{1-\cos 2\theta }{\sin ^22\alpha \sin ^22\theta \sin 2\phi }.
\end{equation}
Furthermore using Eq.\thinspace (A-82), Eq.\thinspace (67) becomes 
\begin{equation}
\label{eq}
\begin{array}{c}
q=\left( 2-\tan ^22\alpha \right) \left[ \cot ^22\alpha -\cos 2\theta \left(
\sin 2\phi +\cot ^22\alpha \right) \right]
\;\;\;\;\;\;\;\;\;\;\;\;\;\;\;\;\;\;\;\;\;\;\;\;\;\;\;\; \\ 
+\sin 2\phi \left[ 1+\left( 1-\tan ^22\alpha \right) \cos 2\theta \right]
-4E\csc ^22\alpha +3.
\end{array}
\end{equation}
Next equating Eqs.\thinspace (A-86) and (A-87) requires 
\begin{equation}
\label{eq}
\begin{array}{c}
\left( 1-2\cot ^22\alpha -\sin 2\phi \right) \sin ^22\theta \sin 2\phi
\;\;\;\;\;\;\;\;\;\;\;\;\;\;\;\;\;\;\;\;\;\;\;\;\;\;\;\; \\ 
=\left( 1-\cos 2\theta \right) \cos 2\theta \cos ^22\phi .
\end{array}
\end{equation}

Next, multiplying Eq.\thinspace (66) by sin$^2\lambda $ and substituting Eq.\thinspace (A-82), one gets 
\begin{equation}
\label{eq}\cos 2\theta =\frac{1-2E}{1-\sin ^22\alpha \left( 1-\sin 2\phi
\right) }. 
\end{equation}
Then substituting Eq.\thinspace (A-90) in Eq.\thinspace (A-89), one obtains 
\begin{equation}
\label{eq}
\begin{array}{c}
\left( 1-2\cot ^22\alpha -\sin 2\phi \right) \sin 2\phi
\;\;\;\;\;\;\;\;\;\;\;\;\;\;\;\;\;\;\;\;\;\;\;\;\;\;\;\;\;\;\;\;\;\;\;\;\;\;
\;\;\;\;\;\;\;\;\;\;\;\;\;\;\;\;\;\;\; \\ 
\times \left\{ \left[ 1-\sin ^22\alpha \left( 1-\sin 2\phi \right) \right]
^2-\left( 1-2E\right) ^2\right\}
\;\;\;\;\;\;\;\;\;\;\;\;\;\;\;\;\;\;\;\;\;\;\;\;\;\;\; \\ 
=(1-2E)\cos ^22\phi \left[ 1-\sin ^22\alpha \left( 1-\sin 2\phi \right)
-(1-2E)\right] , 
\end{array}
\end{equation}
or equivalently, 
\begin{equation}
\label{eq}
\begin{array}{c}
\left[ 1-\sin ^22\alpha \left( 1-\sin 2\phi \right) -(1-2E)\right] \left\{
(1-2E)\cos ^22\phi \right.
\;\;\;\;\;\;\;\;\;\;\;\;\;\;\;\;\;\;\;\;\;\;\;\;\;\;\;\;\;\;\; \\ 
-\left. \left( 1-2\cot ^22\alpha -\sin 2\phi \right) \sin 2\phi \left[
1-\sin ^22\alpha \left( 1-\sin 2\phi \right) +(1-2E)\right] \right\} =0. 
\end{array}
\end{equation}
Therefore, either 
\begin{equation}
\label{eq}\left[ 1-\sin ^22\alpha \left( 1-\sin 2\phi \right) -(1-2E)\right]
=0, 
\end{equation}
or else, 
\begin{equation}
\label{eq}
\begin{array}{c}
(1-2E)\cos ^22\phi -\left( 1-2\cot ^22\alpha -\sin 2\phi \right) \sin 2\phi
\\ 
\;\;\;\;\;\times \left[ 1-\sin ^22\alpha \left( 1-\sin 2\phi \right)
+(1-2E)\right] =0. 
\end{array}
\end{equation}
Equation (A-93) gives 
\begin{equation}
\label{eq}\sin 2\phi =1-2E\csc ^22\alpha , 
\end{equation}
which when substituted in Eq.\thinspace (A-90) gives 
\begin{equation}
\label{eq}\cos 2\theta =1, 
\end{equation}
and substituting Eqs.\thinspace (A-82), (A-95), (A-96) and (8)--(11) in Eq.\thinspace (62), one again obtains
the same solution resulting from possibility (B), Eqs.\thinspace (A-55)--(A-60). However Eq.\thinspace (A-95)
must also be compatible with the remaining requirements if possibility (D) is to represent a solution. 

Alternatively, one has Eq.\thinspace (A-94), which becomes the cubic: 
\begin{equation}
\label{eq}a_1\sin ^32\phi +a_2\sin ^22\phi +a_3\sin 2\phi +a_4=0, 
\end{equation}
where 
\begin{equation}
\label{eq}a_1=\sin ^22\alpha , 
\end{equation}
\begin{equation}
\label{eq}a_2=3-4\sin ^22\alpha , 
\end{equation}
\begin{equation}
\label{eq}a_3=\left( 2E-\cos ^22\alpha -1\right) \left( 1-2\cot ^22\alpha
\right) , 
\end{equation}
\begin{equation}
\label{eq}a_4=(1-2E). 
\end{equation}
The possible solutions to the cubic Eq.\thinspace (A-97) are given by 
\begin{equation}
\label{eq}\sin 2\phi =x_{}-\frac p3, 
\end{equation}
\begin{equation}
\label{eq}\sin 2\phi =x_{+}-\frac p3, 
\end{equation}
\begin{equation}
\label{eq}\sin 2\phi =x_{-}-\frac p3, 
\end{equation}
where 
\begin{equation}
\label{eq}x=c_{+}+c_{-}, 
\end{equation}
\begin{equation}
\label{eq}x_{\pm }=-\frac 12\left( c_{+}+c_{-}\right) \pm \frac{3^{1/2}}
2i\left( c_{+}-c_{-}\right) , 
\end{equation}
\begin{equation}
\label{eq}c_{\pm }=\left[ -\frac B2\pm \left( \frac{B^2}4+\frac{A^3}{27}
\right) ^{1/2}\right] ^{1/3}, 
\end{equation}
\begin{equation}
\label{eq}A=\frac 13\left( 3q-p^2\right) , 
\end{equation}
\begin{equation}
\label{eq}B=\frac 1{27}\left( 2p^3-9pq+27r\right) , 
\end{equation}
\begin{equation}
\label{eq}p=\frac{a_2}{a_1}, 
\end{equation}
\begin{equation}
\label{eq}q=\frac{a_3}{a_1}, 
\end{equation}
\begin{equation}
\label{eq}r=\frac{a_4}{a_1}. 
\end{equation}

Next, substituting Eqs.\thinspace (A-85) and (A-90) in Eq.\thinspace (A-87), one obtains 
\begin{equation}
\label{eq}
\begin{array}{c}
\left[ 2E-\sin ^22\alpha \left( 1-\sin 2\phi \right) \right] \left[ 4\left(
1-E\right) ^2-
\frac{\left[ 2(1-E)-\sin ^22\alpha \left( 1-\sin 2\phi \right) \right] }{
\left[ 1-\sin ^22\alpha \left( 1-\sin 2\phi \right) \right] ^2}\right.
\;\;\;\;\;\;\;\;\;\;\;\;\;\; \\ \;\;\;\;\;\;\times \left\{ \left[ 2E-\sin
^22\alpha \left( 1-\sin 2\phi \right) \right] \sin ^22\alpha \cos ^22\phi
+\left[ 1-\sin ^22\alpha \left( 1-\sin 2\phi \right) \right] \right.  \\ 
\times \left. \left. \left( q-1+2E\right) \sin ^22\alpha \sin 2\phi \right\}
\right] =0.
\end{array}
\end{equation}
Therefore, either 
\begin{equation}
\label{eq}\left[ 2E-\sin ^22\alpha \left( 1-\sin 2\phi \right) \right] =0,
\end{equation}
or else, 
\begin{equation}
\label{eq}
\begin{array}{c}
4\left( 1-E\right) ^2=
\frac{\left[ 2(1-E)-\sin ^22\alpha \left( 1-\sin 2\phi \right) \right] }{
\left[ 1-\sin ^22\alpha \left( 1-\sin 2\phi \right) \right] ^2}
\;\;\;\;\;\;\;\;\;\;\;\;\;\;\;\;\;\;\;\;\;\;\;\;\;\;\;\;\;\;\;\;\;\;\;\;\;\;
\;\;\;\;\;\; \\ \;\;\;\;\;\;\;\;\;\times \left\{ \left[ 2E-\sin ^22\alpha
\left( 1-\sin 2\phi \right) \right] \sin ^22\alpha \cos ^22\phi +\left[
1-\sin ^22\alpha \left( 1-\sin 2\phi \right) \right] \right.  \\ 
\times \left. \left( q-1+2E\right) \sin ^22\alpha \sin 2\phi \right\} .
\end{array}
\end{equation}
Equation (A-114) gives 
\begin{equation}
\label{eq}\sin 2\phi =1-2E\csc ^22\alpha ,
\end{equation}
which together with Eqs.\thinspace (A-90), (A-82), (8)--(11), and (62) again yields the same result as
possibility (B), Eqs.\thinspace (A-55)--(A-60). However, Eqs.\thinspace (A-95) and (A-116) must also be
compatible with the remaining restrictions, if possibility (D) is to represent a solution. 

Alternatively, one has Eq.\thinspace (A-115). The quantity $q$ appearing in Eq.\thinspace (A-115) and given by
Eq.\thinspace (A-88) reduces using Eq.\thinspace (A-90) to  
\begin{equation}
\label{eq}q=\sin 2\phi +\frac{\left( 1+\sin 2\phi \right) \left( 1-2E\right) 
}{\cos ^22\alpha +\sin ^22\alpha \sin 2\phi }.
\end{equation}
Then substituting Eq.\thinspace (A-117) in Eq.\thinspace (A-115), one obtains the cubic: 
\begin{equation}
\label{eq}b_1\Lambda ^3+b_2\Lambda ^2+b_3\Lambda +b_4=0,
\end{equation}
where 
\begin{equation}
\label{eq}\Lambda =\cos ^22\alpha +\sin ^22\alpha \sin 2\phi ,
\end{equation}
\begin{equation}
\label{eq}b_1=(1-2E)\left( 1-2\csc ^22\alpha \right) ,
\end{equation}
\begin{equation}
\label{eq}
\begin{array}{c}
b_2=4(1-E)^2-\sin ^22\alpha +(1-2E)^2\left( 1-2\csc ^22\alpha \right)
\;\;\;\;\;\;\;\;\;\;\;\;\;\; \\ 
-(1-2E)\left( 1+\cos ^22\alpha -4\cot ^22\alpha \right) ,
\end{array}
\end{equation}
\begin{equation}
\label{eq}b_3=-(1-2E)^2\left( 1+\cos ^22\alpha -4\cot ^22\alpha \right)
+(1-2E)\cos ^22\alpha \left( 1-2\cot ^22\alpha \right) ,
\end{equation}
\begin{equation}
\label{eq}b_4=(1-2E)^2\left( 1-2\cos ^22\alpha \cot ^22\alpha \right) .
\end{equation}
(In obtaining Eq.\thinspace (A-118), an overall factor of $\Lambda $ was removed and ignored, since $\Lambda
=0$ can only be satisfied if $E=1/2$.) 

Next, substituting Eqs.\thinspace (A-85), (A-90) and (A-117) in Eq.\thinspace (A-86), leads to the quintic: 
\begin{equation}
\label{eq}c_1\sin ^52\phi +c_2\sin ^42\phi +c_3\sin ^32\phi +c_4\sin ^22\phi
+c_5\sin 2\phi +c_6=0, 
\end{equation}
where 
\begin{equation}
\label{eq}c_1=\sin ^62\alpha , 
\end{equation}
\begin{equation}
\label{eq}c_2=\sin ^42\alpha \left( 5\cos ^22\alpha +2E-2\right) , 
\end{equation}
\begin{equation}
\label{eq}
\begin{array}{c}
c_3=\sin ^42\alpha \left( 5-12E+8E^2\right) -\sin ^22\alpha \cos ^22\alpha
(1-2E)-2\sin ^22\alpha \left( 1-2E\right) ^2\; \\ 
\;\;\;\;\;\;\;\;\;\;\;-2\sin ^42\alpha \cos ^22\alpha +5\sin ^22\alpha \cos
^42\alpha -\sin ^62\alpha , 
\end{array}
\end{equation}
\begin{equation}
\label{eq}
\begin{array}{c}
c_4=\left( 1-2\cot ^22\alpha \right) \left[ \sin ^22\alpha (1-2E)^2-4\sin
^42\alpha (1-E)^2+\sin ^62\alpha \right.
\;\;\;\;\;\;\;\;\;\;\;\;\;\;\;\;\;\;\;\;\;\; \\ 
\;-\left. \sin ^22\alpha \cos ^42\alpha \right] -2\sin ^42\alpha \cos
^22\alpha -\sin ^42\alpha (1-2E)^2 \\ 
+\sin ^42\alpha \left( 1-2E\right) +8\sin ^22\alpha \cos ^22\alpha \left(
1-E\right) ^2, 
\end{array}
\end{equation}
\begin{equation}
\label{eq}
\begin{array}{c}
c_5=\left( 1-2\cot ^22\alpha \right) \left[ -8\sin ^22\alpha \cos ^22\alpha
(1-E)^2+2\sin ^42\alpha \cos ^22\alpha \right] \; \\ 
\;\;\;\;\;\;\;\;\;\;+4\cos ^42\alpha \left( 1-E\right) ^2+\sin ^22\alpha
\left( 2-\sin ^22\alpha \right) \left( 1-2E\right) ^2 \\ 
\;\;\;\;\;\;\;\;\;+\sin ^22\alpha \cos ^22\alpha (1-2E)-\sin ^22\alpha \cos
^42\alpha , 
\end{array}
\end{equation}
\begin{equation}
\label{eq}
\begin{array}{c}
c_6=\left( 1-2\cot ^22\alpha \right) \left[ \sin ^22\alpha \cos ^42\alpha
-4\cos ^42\alpha (1-E)^2-\sin ^22\alpha (1-2E)^2\right] \; \\ 
\;\;\;\;\;\;\;\;\;\;+\sin ^42\alpha (1-2E)^2. 
\end{array}
\end{equation}

In summary, the possibility (D) requires that one of the following three sets of equations be satisfied: 
\begin{equation}
\label{eq}(i)\;\;\text{Eqs.\thinspace (A-97), (A-118), and (A-124);} 
\end{equation}
\begin{equation}
\label{eq}(ii)\;\;\text{Eqs.\thinspace (A-95), (A-118), and (A-124);} 
\end{equation}
\begin{equation}
\label{eq}(iii)\;\;\text{Eqs.\thinspace (A-95) and (A-124).} 
\end{equation}
But none of these alternatives, (i), (ii), or (iii) can be satisfied. It can be shown numerically that
Eqs.\thinspace (A-97), (A-118), and (A-124) cannot be simultaneously satisfied. Evidently, it can also be shown
numerically that Eqs.\thinspace (A-95) and (A-124) cannot be simultaneously satisfied. (This has been
verified explicitly for $\alpha =\pi /9$, $\pi /8$, and $\pi /5$.) Thus, possibility (D) apparently does
not produce a solution. 

Next, consider possibility (E), given by Eq.\thinspace (A-43): 
\begin{equation}
\label{eq}\cos \lambda =0. 
\end{equation}
Substituting Eq.\thinspace (A-134) in Eq.\thinspace (66), one has 
\begin{equation}
\label{eq}\sin 2\mu =1-2E\csc ^22\alpha . 
\end{equation}
Next substituting Eqs.\thinspace (A-134) and (A-135) in Eqs.\thinspace (8)--(11), one obtains 
\begin{equation}
\label{eq}a=1-2E\csc ^22\alpha , 
\end{equation}
\begin{equation}
\label{eq}b=1-2E\csc ^22\alpha , 
\end{equation}
\begin{equation}
\label{eq}c=0, 
\end{equation}
\begin{equation}
\label{eq}d=1. 
\end{equation}
Then substituting Eqs.\thinspace (A-136)--(A-139) in Eq.\thinspace (62), one again obtains Eq.\thinspace
(A-60). Therefore, possibility (E), Eq.\thinspace (A-43), gives the same result as possibility (B),
Eq.\thinspace (A-40). Note however that the probe parameter $\mu $ is restricted by Eq.\thinspace (A-135),
and the probe parameter $\phi $ is unrestricted, while for possibility (B), $\phi $ is restricted by
Eq.\thinspace (A-55), and $\mu $ is unrestricted. This is addressed in Section 4. 

Next, consider possibility (F), given by Eq.\thinspace (A-44): 
\begin{equation}
\label{eq}\sin 2\theta =0, 
\end{equation}
\begin{equation}
\label{eq}\cos 2\theta =e_\theta , 
\end{equation}
\begin{equation}
\label{eq}\cos 2\phi =0, 
\end{equation}
\begin{equation}
\label{eq}\sin 2\phi =e_\phi , 
\end{equation}
\begin{equation}
\label{eq}F_1=0. 
\end{equation}
Substituting Eqs.\thinspace (A-5), and (A-140)--(A-143) in Eq.\thinspace (A-144), one requires 
\begin{equation}
\label{eq}\left( 1-e_\theta \right) \left[ e_\phi \cot ^22\alpha \left(
2-\tan ^22\alpha \right) +1\right] =0, 
\end{equation}
and therefore 
\begin{equation}
\label{eq}e_\theta =1. 
\end{equation}
Next substituting Eqs.\thinspace (A-141), (A-146), and (A-143) in Eq.\thinspace (66), one gets 
\begin{equation}
\label{eq}\sin 2\mu =\frac{\sin ^22\alpha \left( 1-e_\phi \cos ^2\lambda
\right) -2E}{\sin ^22\alpha \sin ^2\lambda }. 
\end{equation}
Then substituting Eqs.\thinspace (A-140)--(A-143), (A-146) and (A-147) in Eqs.\thinspace (8)--(11), one again
obtains Eqs.\thinspace (A-136)--(A-139), and (A-60). Thus possibility (F), Eq.\thinspace (A-44), also gives the
same result as possibility (B), Eq.\thinspace (A-40). Note however that the probe parameters $\mu $ and
$\lambda $ are restricted by Eq.\thinspace (A-147). This is addressed in Section 4. 

Next consider possibility (G), given by Eq.\thinspace (A-45): 
\begin{equation}
\label{eq}\cos \lambda =0, 
\end{equation}
\begin{equation}
\label{eq}\sin 2\theta =0 
\end{equation}
\begin{equation}
\label{eq}\cos 2\theta =e_\theta , 
\end{equation}
\begin{equation}
\label{eq}F_1=0. 
\end{equation}
Substituting Eq.\thinspace (A-148) in Eq.\thinspace (66), one gets 
\begin{equation}
\label{eq}\sin 2\mu =1-2E\csc ^22\alpha . 
\end{equation}
Substituting Eqs.\thinspace (A-5) and (A-148)--(A-150) in Eq.\thinspace (A-151), one obtains 
\begin{equation}
\label{eq}\left( 1-e_\theta \right) \left[ \sin 2\phi +\cot ^22\alpha \left(
2-\tan ^22\alpha \right) \right] =0. 
\end{equation}
Therefore, one requires 
\begin{equation}
\label{eq}e_\theta =1, 
\end{equation}
or alternatively, 
\begin{equation}
\label{eq}\sin 2\phi =1-2\cot ^22\alpha . 
\end{equation}
Substituting Eqs.\thinspace (A-148), (A-149), and (A-152) in Eqs.\thinspace (8)--(11), one again obtains
Eqs.\thinspace (A-136)--(A-139) and (A-60). The differing values of the probe parameters are addressed in
Section 4. 

Next consider possibility (H), given by Eq.\thinspace (A-46): 
\begin{equation}
\label{eq}\sin 2\theta =0 
\end{equation}
\begin{equation}
\label{eq}\cos 2\theta =e_\theta , 
\end{equation}
\begin{equation}
\label{eq}F_1=0, 
\end{equation}
\begin{equation}
\label{eq}F_3=0. 
\end{equation}
Substituting Eqs.\thinspace (A-15) and (A-156) in Eq.\thinspace (A-159), one gets 
\begin{equation}
\label{eq}\cos 2\theta =1, 
\end{equation}
and therefore 
\begin{equation}
\label{eq}e_\theta =1 
\end{equation}
in Eq.\thinspace (A-157). Next using Eqs.\thinspace (A-5) and (A-160), one sees that Eq.\thinspace (A-158) is
satisfied. Also, substituting Eq.\thinspace (A-160) in Eq.\thinspace (66), one obtains  
\begin{equation}
\label{eq}\sin 2\mu =\frac{\sin ^22\alpha \left( 1-\cos ^2\lambda \sin 2\phi
\right) -2E}{\sin ^22\alpha \sin ^2\lambda }. 
\end{equation}
Then substituting Eqs.\thinspace (A-156), (A-160), and (A-162) in Eqs.\thinspace (8)--(11), one again obtains
Eqs.\thinspace (A-136)--(A-139) and (A-60). The differing values of the probe parameters are addressed in
Section 4. 

Next consider possibility (I), given by Eq.\thinspace (A-47): 
\begin{equation}
\label{eq}\cos \lambda =0 
\end{equation}
\begin{equation}
\label{eq}F_1=0, 
\end{equation}
Substituting Eqs.\thinspace (A-163) in Eq.\thinspace (66), one gets 
\begin{equation}
\label{eq}\sin 2\mu =1-2E\csc ^22\alpha . 
\end{equation}
Next substituting Eqs.\thinspace (A-5) and (A-163) in Eq.\thinspace (A-164), one obtains 
\begin{equation}
\label{eq}\left( 1-\cos 2\theta \right) \left[ \sin 2\phi +2\cot ^22\alpha
-1\right] =0. 
\end{equation}
Therefore one requires 
\begin{equation}
\label{eq}\cos 2\theta =1, 
\end{equation}
or else, 
\begin{equation}
\label{eq}\sin 2\phi =1-2\cot ^22\alpha . 
\end{equation}
Using Eqs.\thinspace (A-163), (A-165), and (A-167) or (A-168) in Eqs.\thinspace (8)--(11), one again obtains
Eqs.\thinspace (A-136)--(A-139) and (A-60). The differing values of the probe parameters are addressed in
Section 4. 

Next consider possibility (J), given by Eqs.\thinspace (A-48): 
\begin{equation}
\label{eq}\cos 2\phi =0, 
\end{equation}
\begin{equation}
\label{eq}\sin 2\phi =e_\phi , 
\end{equation}
\begin{equation}
\label{eq}F_1=0, 
\end{equation}
\begin{equation}
\label{eq}F_2=0. 
\end{equation}
Then substituting Eqs.\thinspace (A-10) and (A-170) in Eq.\thinspace (A-172), one gets 
\begin{equation}
\label{eq}\cot ^22\alpha =\frac 12\left( 1-e_\phi \right) , 
\end{equation}
which cannot be satisfied for arbitrary $\alpha .$ Therefore possibility (J) cannot represent a solution for
arbitrary $\alpha$. (It is to be noted however that Eq. (A-173) is satisfied if $e_\rho=-1$, and $\alpha=\pi/8$.
This particular case is addressed in Section 6, following Eq. (119).)

Next consider possibility (K), given by Eqs.\thinspace (A-49): 
\begin{equation}
\label{eq}\cos \lambda =0,
\end{equation}
\begin{equation}
\label{eq}F_1=0,
\end{equation}
\begin{equation}
\label{eq}F_2=0.
\end{equation}
Substituting Eqs.\thinspace (A-10) and (A-174) in Eq.\thinspace (A-176), one obtains 
\begin{equation}
\label{eq}\sin 2\phi =1-2\cot ^22\alpha .
\end{equation}
Next substituting Eqs.\thinspace (A-5), (A-174), and (A-177) in Eq.\thinspace (A-175), one gets the trivial
identity:  
\begin{equation}
\label{eq}
\begin{array}{c}
\left( 2-\tan ^22\alpha \right) \left[ \cot ^22\alpha -\cos 2\theta \left(
1-\cot ^22\alpha \right) \right]
\;\;\;\;\;\;\;\;\;\;\;\;\;\;\;\;\;\;\;\;\;\;\;\;\;\;\;\;\;\;\;\;\;\;\;\;\;\;
\;\;\;\; \\ 
+\left( 1-2\cot ^22\alpha \right) \left[ 1+\left( 1-\tan ^22\alpha \right)
\cos 2\theta \right] =0
\end{array}
\end{equation}
for any $\cos 2\theta $. Then substituting Eq.\thinspace (A-174) in Eq.\thinspace (66), one obtains 
\begin{equation}
\label{eq}\sin 2\mu =1-2E\csc ^22\alpha ,
\end{equation}
and, using Eqs.\thinspace (A-174), (A-179), (8)--(11), and (62), then Eqs.\thinspace (A-136)--(A-139) and (A-60)
again follow. The differing values of the probe parameters are addressed in Section 4. 

Next consider possibility (L), given by Eqs.\thinspace (A-50): 
\begin{equation}
\label{eq}F_1=0, 
\end{equation}
\begin{equation}
\label{eq}F_2=0, 
\end{equation}
\begin{equation}
\label{eq}F_3=0. 
\end{equation}
From Eqs.\thinspace (A-5) and (A-180), it follows that 
\begin{equation}
\label{eq}
\begin{array}{c}
\sin ^22\alpha \cos ^2\lambda \left[ 
\frac{2\left( q-1+2E\right) }{4\left( 1-E\right) ^2-c^2\sin ^22\alpha }
\right]
\;\;\;\;\;\;\;\;\;\;\;\;\;\;\;\;\;\;\;\;\;\;\;\;\;\;\;\;\;\;\;\;\;\;\;\;\;\;
\;\;\;\;\;\;\;\;\;\;\;\;\; \\ =\frac{-2\left\{ \left( 2-\tan ^22\alpha
\right) \left[ \cot ^22\alpha -\cos 2\theta \left( \sin 2\phi +\cot
^22\alpha \right) \right] +\sin 2\phi \left[ 1+\left( 1-\tan ^22\alpha
\right) \cos 2\theta \right] \right\} }{\sin ^22\theta \cos ^22\phi }. 
\end{array}
\end{equation}
From Eqs.\thinspace (A-10) and (A-181), one gets 
\begin{equation}
\label{eq}\sin ^22\alpha \cos ^2\lambda \left[ \frac{2\left( q-1+2E\right) }{
4\left( 1-E\right) ^2-c^2\sin ^22\alpha }\right] =\frac{-2\left( \sin 2\phi
+2\cot ^22\alpha -1\right) }{\cos 2\theta \cos ^22\phi }. 
\end{equation}
From Eqs.\thinspace (A-15) and (A-182), one gets 
\begin{equation}
\label{eq}\sin ^22\alpha \cos ^2\lambda \left[ \frac{2\left( q-1+2E\right) }{
4\left( 1-E\right) ^2-c^2\sin ^22\alpha }\right] =\frac{2\left( 1-\cos
2\theta \right) }{\sin ^22\theta \sin 2\phi }. 
\end{equation}
Next equating Eqs.\thinspace (A-183) and (A-185) leads to 
\begin{equation}
\label{eq}\cos 2\theta =1, 
\end{equation}
and Eqs.\thinspace (A-183) and (A-185) are both identically satisfied. But then substituting Eq.\thinspace
(A-186), (67), and (10) in Eq.\thinspace (A-184), one obtains  
\begin{equation}
\label{eq}\sin ^22\alpha \cos ^2\lambda \cos ^22\phi \left[ 1+\left( 1-2\csc
^22\alpha \right) E\right] =-2\left( 1-E\right) ^2\left( \sin 2\phi +2\cot
^22\alpha -1\right) , 
\end{equation}
or 
\begin{equation}
\label{eq}\cos ^2\lambda =\frac{2\left( 1-E\right) ^2\left( 1-2\cot
^22\alpha -\sin 2\phi \right) }{\sin ^22\alpha \cos ^22\phi \left[ 1+\left(
1-2\csc ^22\alpha \right) E\right] }. 
\end{equation}
Also, substituting Eq.\thinspace (A-186) in Eq.\thinspace (66), one obtains 
\begin{equation}
\label{eq}\sin 2\mu \sin ^2\lambda =1-2E\csc ^22\alpha -\cos ^2\lambda \sin
2\phi . 
\end{equation}
Then substituting Eqs.\thinspace (A-188) and (A-189) in Eqs.\thinspace (8)--(11) yields 
\begin{equation}
\label{eq}a=1-2E\csc ^22\alpha , 
\end{equation}
\begin{equation}
\label{eq}b=1-2E\csc ^22\alpha , 
\end{equation}
\begin{equation}
\label{eq}c=0, 
\end{equation}
\begin{equation}
\label{eq}d=1. 
\end{equation}
Next substituting Eqs.\thinspace (A-190)--(A-193) in Eq.\thinspace (62) again leads to 
\begin{equation}
\label{eq}Q=\frac{1+\left( 1-2\csc ^22\alpha \right) E}{1-E}. 
\end{equation}
The differing values of the probe parameters are addressed in Section 4.\\
REFERENCES

[1] N. Gisin, G. Ribordy, W. Tittel, and H. Zbinden, ``Quantum
cryptography,'' Rev. Mod. Phys. {\bf 74}, 145--195 (2002). 

[2] S. Wiesner, ``Conjugate coding,'' SIGACT News {\bf 15}, 78--88 (1983).

[3] C. H. Bennett and G. Brassard, ``Quantum cryptography: public key distribution and coin tossing,'' in
Proceedings of the IEEE International Conference on Computers, Systems, and Signal Processing, Bangalore,
India (IEEE, New York, 1984), pp. 175--179. 

[4] C. H. Bennett and G. Brassard, ``Quantum public key distribution system,'' IBM Tech. Discl. Bull. {\bf
28}, 3153--3163 (1985). 

[5] G. Vernam, ``Cipher printing telegraph systems for secret wire and radio telegraphic communication, ''
J. Am. Inst. Electr. Eng. {\bf 45}, 295--301 (1926). 

[6] C. H. Bennett,``Quantum cryptography using any two nonorthogonal states,'' Phys. Rev. Lett. {\bf 68},
3121--3124 (1992). 

[7] A. K. Ekert, ``Quantum cryptography based on Bell's theorem,'' Phys. Rev. Lett. {\bf 67}, 661--663
(1991). 

[8] H. E. Brandt, ``Positive operator valued measure in quantum information processing,'' Am. J. Phys.
{\bf 67}, 434--439 (1999). 

[9] C. H. Bennett, G. Brassard, C. Crepeau, and U. M. Maurer, ``Generalized privacy amplification,'' IEEE
Trans. Inf. Theor. {\bf 41}, 1915--1923 (1995). 

[10] C. H. Bennett, F. Bessette, G. Brassard, L. Salvail, and J. Smolin, ``Experimental quantum
cryptography,'' J. Cryptology {\bf 5}, 3--28 (1992). 

[11] B. A. Slutsky, R. Rao, P. C. Sun, and Y. Fainman, ``Security of quantum cryptography against
individual attacks,'' Phys. Rev. A {\bf 57}, 2383--2398 (1998). 

[12] H. E. Brandt, ``Probe Optimization in four-state protocol of quantum cryptography,'' Phys. Rev. A
{\bf 66}, 032303-1-16 (2002). 

[13] H. E. Brandt, ``Secrecy capacity in the four-state protocol of quantum key distribution,'' J.
Math. Phys. {\bf 43}, 4526-4530 (2002). 

[14] H. E. Brandt, ``Optimization Problem in Quantum Cryptography,'' to appear in J. Opt. B
(2003). 

[15] C. A. Fuchs and A. Peres, ``Quantum-state disturbance versus information gain: uncertainty relations
for quantum information,'' Phys. Rev. A {\bf 53}, 2038--2045 (1996). 

[16] H. E. Brandt, ``Eavesdropping optimization for quantum cryptography using a positive operator valued
measure,'' Phys. Rev. A {\bf 59},  2665--2669 (1999).

[17] H. E. Brandt, ``Inconclusive rate as a disturbance measure in quantum cryptography,'' Phys. Rev. A
{\bf 62}, 042310-1-14 (2000). 

[18] H. E. Brandt, ``Inconclusive rate in quantum key distribution,'' Phys. Rev. A {\bf 64}, 042316-1-5
(2001). 

[19] B. Slutsky, R. Rao, P. C. Sun, L. Tancevski, and S. Fainman, ``Defense frontier analysis of quantum
cryptographic systems,'' Applied Optics {\bf 37}, 2869--2878 (1998).

\end{document}